\documentclass[journal]{IEEEtran}

\IEEEoverridecommandlockouts
\usepackage{amsmath,amssymb,amsfonts}
\usepackage[utf8]{inputenc}
\usepackage{algorithm}
\usepackage[noend]{algpseudocode}
\algrenewcommand\algorithmicrequire{\textbf{Input:}}
\algrenewcommand\algorithmicensure{\textbf{Output:}}
\usepackage{array}
\usepackage{textcomp}
\usepackage{url}
\usepackage{graphics}
\usepackage{graphicx}
\usepackage{verbatim}
\usepackage{cite} 
\usepackage{siunitx}
\usepackage{booktabs, makecell}

\ifCLASSOPTIONcompsoc
    \usepackage[caption=false, font=normalsize, labelfont=sf, textfont=sf]{subfig}
\else
    \usepackage[caption=false, font=footnotesize]{subfig}
\fi
\usepackage{nomencl}
\usepackage{multirow,tabularx}
\usepackage[table]{xcolor}
\usepackage{etoolbox}
\usepackage{array,multirow}
\usepackage{colortbl}
\usepackage{soul}
\sethlcolor{yellow}
\usepackage{mathtools}
\usepackage{mdwtab}
\makeatletter
\newcommand*{\rom}[1]{\expandafter\@slowromancap\romannumeral #1@}
\makeatother
\usepackage{float}
\graphicspath{{./Figures/}}
\usepackage{comment}
\usepackage{gensymb} 
\usepackage{bm} 
\usepackage{cleveref} 
\makeatletter
\newcommand{\linebreakand}{%
  \end{@IEEEauthorhalign}
  \hfill\mbox{}\par
  \mbox{}\hfill\begin{@IEEEauthorhalign}
}
\makeatother

\begin{document}

\title{An Iterative Bidirectional Gradient Boosting Approach for CVR Baseline Estimation\\}

\author{Han~Pyo~Lee,~\IEEEmembership{Student Member,~IEEE}, Yiyan~Li,~\IEEEmembership{Member,~IEEE},\\ Lidong~Song,~\IEEEmembership{Member,~IEEE}, Di~Wu,~\IEEEmembership{Senior Member,~IEEE}, Ning~Lu,~\IEEEmembership{Fellow,~IEEE} \vspace{-0.4in} \thanks{This research is supported by the U.S. Department of Energy's Office of Energy Efficiency and Renewable Energy (EERE) under the Solar Energy Technologies Office Award Number DE-EE0008770. Han~Pyo~Lee, Lidong~Song, and Ning~Lu are with the Electrical \& Computer Engineering Department, Future Renewable Energy Delivery and Management (FREEDM) Systems Center, North Carolina State University, Raleigh, NC 27606 USA, (e-mails: \{hlee39, lsong4, nlu2\}@ncsu.edu). Yiyan~Li is with the College of Smart Energy, Shanghai Jiao Tong University, Shanghai, 200240, China, (e-mail: yiyan.li@sjtu.edu.cn). Di~Wu is with Pacific Northwest National Laboratory, (e-mail: Di.Wu@pnnl.gov).}}
\maketitle

\bstctlcite{IEEEexample:BSTcontrol}
\begin{abstract}
This paper presents a novel Iterative Bidirectional Gradient Boosting Model (IBi-GBM) for estimating the baseline of Conservation Voltage Reduction (CVR) programs. In contrast to many existing methods, we treat CVR baseline estimation as a missing data retrieval problem. The approach involves dividing the load and its corresponding temperature profiles into three periods: pre-CVR, CVR, and post-CVR. To restore the missing load profile during the CVR period, the method employs a three-step process. First, a forward-pass GBM is executed using data from the pre-CVR period as inputs. Subsequently, a backward-pass GBM is applied using data from the post-CVR period. The two restored load profiles are reconciled, considering pre-calculated weights derived from forecasting accuracy, and only the leftmost and rightmost points are retained. The newly restored points are then included as inputs for the subsequent iteration. This iterative procedure continues until the original load data in the CVR period is fully restored. We develop IBi-GBM using actual smart meter and Supervisory Control and Data Acquisition (SCADA) data. Our results demonstrate that IBi-GBM exhibits robust performance across various data resolutions and in different seasons and outperforms existing methods by achieving a 1-2\% reduction in normalized Root Mean Square Error (nRMSE).
\end{abstract}


\vspace{-0.1in}
\section{Introduction}
\IEEEPARstart{A}{ccurate} baseline estimation is crucial for utilities to assess the effectiveness of demand response (DR) programs \cite{IEEECVRreport}. The baseline of a DR event is essentially the load profile that would exist if a DR action had not been implemented. Thus, to estimate the DR baseline, it's essential to reconstruct the electricity consumption as it would be during the DR period, assuming that no DR actions are implemented. This process can be conceptualized as a missing data segments (MSDs) recovery problem, as detailed in \cite{li2023load}.

For instance, in a CVR event, the voltage provided to the distribution feeder is deliberately lowered within a 2-5\% range to attain load reduction. The duration of a CVR event may vary, depending on the anticipated duration of the system peak load. As the voltage is at the same level in both pre-CVR and post-CVR periods, computing the CVR baseline is akin to restoring the missing data points in the on-CVR period.

In Table~\ref{tab1}, we offer a thorough literature review of current CVR baseline estimation approaches, highlighting their strengths and weaknesses in comparison to the proposed approach. In general, there are five main approaches for CVR baseline estimation: comparison-based, synthesis-based, load modeling-based, regression-based, and machine learning (ML)-based techniques. 

\textit{Comparison-based} methods represent an experimental approach where the load consumption of the voltage-reduction group (test group) and the normal voltage group (control group)  \cite{kennedy1991conservation} is compared through field experiments. The control groups can be the same feeder on a day without voltage reduction or a different feeder with similar operating conditions. The challenge of the comparison-based method is that there may not be an appropriate control group because there are no two feeders or two days with an exact match of operating conditions. 

\textit{Synthesis-based} methods aggregate load-to-voltage (LTV) behaviors to estimate the CVR effects of a circuit \cite{kirshner1990implementation}. Yet, the load composition of load groups exhibiting similar LTV behaviors fluctuates throughout the day, making it an unknown variable for utility engineers, and thereby rendering this method impractical in the field. \textit{Load modeling-based} methods take a similar approach by representing the feeder load consumption as a function of voltages. The CVR factors are calculated from the identified LTV sensitivities \cite{wang2014time, zhao2016robust}. However, it also encounters similar challenges as the synthesis-based method. Due to the changing load composition of groups with similar LTV behaviors throughout the day, the load model parameters can undergo significant shifts in response to ambient temperature fluctuations and changes in customer usage patterns.

\textit{Regression-based} methods employ a linearized model to calculate the load, taking inputs such as temperature, voltage, and other impact factors into account \cite{wilson2010measurement, dwyer1995load}. The problems with this method are that the margin of error of the regression model may be larger than CVR effect, and most existing linear models cannot accurately capture the characteristics of nonlinear loads. 

\begin{table*}[ht]
    \begin{center}
    \caption{A Review of Existing Methods and Our Contributions.}
    \vspace{-0.2in}
    \label{tab1}  
    \resizebox{\linewidth}{!}{%
    \begin{tabularx}{1.02\linewidth}{cllll}
    \toprule
    \textbf{Category} &\thead{Methodology} &\thead{Description} &\thead{Strength} &\thead{Weakness} \\
    \midrule \midrule
    \multirow{2}{5em}{\parbox{1\linewidth}{\centering Comparison \\ -based}}
    &\multirow{2}{10em}{\parbox{1\linewidth}{Field experiments \\ \cite{kennedy1991conservation}}}
    &\multirow{2}{13em}{\parbox{1\linewidth}{Compare performance between test and control groups}}
    &\multirow{2}{14em}{\parbox{1\linewidth}{Simplicity}} 
    &\multirow{2}{16em}{\parbox{1\linewidth}{Dependent on control group}} \\ \\
    \midrule       
    \multirow{4}{5em}{\parbox{1\linewidth}{\centering Synthesis \\ -based}}
    &\multirow{4}{10em}{\parbox{1\linewidth}{LTV \cite{kirshner1990implementation}}}
    &\multirow{4}{13em}{\parbox{1\linewidth}{Aggregate LTV behaviors to estimate the CVR effects of a circuit}}
    &\multirow{4}{14em}{\vspace{-0.1cm} Quick estimation} 
    &\multirow{4}{16em}{\parbox{1\linewidth}{Difficult to collect load share information for a feeder \\ Requires LTV response of all existing electrical appliances}} \\ \\ \\ \\
    \midrule
    \multirow{4}{5em}{\parbox{1\linewidth}{\centering Load \\ modeling \\ -based}}
    &\multirow{4}{10em}{\parbox{1\linewidth}{LTV sensitivity \\ \cite{wang2014time, zhao2016robust}}}
    &\multirow{4}{13em}{\parbox{1\linewidth}{Represent load consumption as a function of voltages, and calculate CVR factors from the identified LTV sensitivities}}
    &\multirow{4}{14em}{Can estimate time varying CVR factors}
    &\multirow{4}{16em}{\parbox{1\linewidth}{Cannot represent different load compositions depending on the load model used}} \\ \\ \\ \\
    \midrule
    \multirow{4}{5em}{\parbox{1\linewidth}{\centering Regression \\ -based}}
    &\multirow{4}{10em}{\parbox{1\linewidth}{Linear regression \\ \cite{wilson2010measurement, dwyer1995load} \\ Multivariate regression \\ \cite{papalexopoulos1990regression, hong2011naive, hong2012impact}}}
    &\multirow{4}{13em}{\parbox{1\linewidth}{Loads are modeled as a function of several impact factors to calculate the CVR factor}}
    &\multirow{4}{14em}{\parbox{1\linewidth}{Interpretable \\ Mathematical simplicity}}
    &\multirow{4}{16em}{\parbox{1\linewidth}{The margin of error may be larger than CVR effect \\ Inability to capture the characteristics of nonlinear loads}} \\ \\ \\ \\
    \midrule
    \multirow{12}{5em}{\parbox{1\linewidth}{\centering ML \\ -based}}
    &\multirow{5}{10em}{MSVR \cite{wang2013analysis}}
    &\multirow{5}{13em}{MSVR-based model}
    &\multirow{5}{14em}{\parbox{1\linewidth}{Interpretable \\ Can approximate nonlinear behaviors of load}}
    &\multirow{5}{16em}{\parbox{1\linewidth}{Noniterative and uni-directional \\ Accuracy relies on the existence of similar profiles \\ Results may not always be attainable for each test day}} \\ \\ \\ \\ \\
    \cmidrule{2-5}
    &\multirow{4}{10em}{MLP, LSTM, TCN \\ \cite{dudek2020multilayer, zheng2017electric, bai2018empirical}}
    &\multirow{4}{13em}{Uni-directional DL model}
    &\multirow{4}{14em}{Can capture complex and nonlinear relationships}
    &\multirow{4}{16em}{\parbox{1\linewidth}{Uni-directional \\ Fixed prediction length \\ Demands a substantial volume of training data}} \\ \\ \\ \\
    \cmidrule{2-5}
    &\multirow{5}{10em}{Load-PIN \cite{li2023load}}
    &\multirow{5}{13em}{GAN-based generative method}
    &\multirow{5}{14em}{\parbox{1\linewidth}{Bi-directional \\ Variable prediction length \\ Can capture complex and nonlinear relationships}}
    &\multirow{5}{16em}{\parbox{1\linewidth}{Mathematically complex \\ Lack of interpretability \\ Computationally expensive for training \\ 
    Demands a substantial volume of training data}} \\ \\ \\ \\ \\
    \midrule
    \multirow{5}{5em}{\parbox{1\linewidth}{\centering \textbf{Iterative} \\ \textbf{bidirectional}}}
    &\multirow{5}{10em}{\parbox{1\linewidth}{\textbf{IBi-GBM} \\ \textbf{IBi-LightGBM} \\ \textbf{(Proposed)}}}
    &\multirow{5}{13em}{\parbox{1\linewidth}{\textbf{Iterative, bidirectional} \\ \textbf{GB-based algorithm}}}
    &\multirow{5}{14em}{\parbox{1\linewidth}{\textbf{Interpretable, capture nonlinear behaviors of load and bi-directional information for variable-duration events and require very few training data}}}
    &\multirow{5}{16em}{\parbox{1\linewidth}{\textbf{Accuracy relies on the existence of available similar profiles ($\geq$ 5 days)}}} \\ \\ \\ \\ \\
    \bottomrule
    \end{tabularx}}
    \end{center}
\vspace{-0.3in}
\end{table*}

Thus, in recent years, \textit{ML-based} methods are used to capture the nonlinear behaviors. Nevertheless, the ML-based approach encounters a significant hurdle due to the scarcity of training data. This is because CVR events typically occur during anticipated system peak periods. Consequently, there are only a few days in a month that can trigger a CVR action. This means there are only tens of days available for training ML algorithms.

Moreover, many deep learning-based approaches (e.g., MLP, LSTM, and TCN) \cite{dudek2020multilayer, zheng2017electric, bai2018empirical} require fixed-length input or output. However, in practice, the duration of CVR events varies based on the duration of anticipated system peak periods. To address the restoration of an MSD with varying duration, our previous work \cite{li2023load} introduced the Load Profile Inpainting Network (Load-PIN), a Generative Adversarial Nets (GAN) model. While Load-PIN excels in restoring MSDs for various load profiles compared to other models, its performance in restoring the CVR baseline can be improved. This is attributed to Load-PIN being trained using load profiles without distinguishing between CVR days and non-CVR days. Consequently, this approach may introduce significant bias, because CVR events usually occur during the system peak load periods.

Lastly, utility engineers often prefer methods that are interpretable. However, many deep learning based methods (e.g., the GAN-based model) lack transparency in terms of how predictions are generated and which factors are most influential for achieving desired amount of load reductions.

In \cite{lee2023iterative}, we introduce a bottom-up method to derive CVR baselines using aggregated smart meter data. The prediction target is power consumption, with a fixed set of weights applied to reconcile forward and backward forecasting results. However, this method has three limitations. First, the bottom-up approach cannot account for line losses. Second, the heuristic approach used in selecting weighting factors is not generalizable to other seasons or other feeders. Third, using power consumption as the prediction target leads to larger errors.

In this paper, we expand our analysis to two approaches: bottom-up (using aggregated smart meters data as inputs) and top-down (using feeder head SCADA data as inputs). We apply the algorithm to data from 15 feeders across two utilities covering both summer and winter seasons. For optimal candidate day selection, we employ both power consumption and temperature as criteria, leading to an average 0.05\% reduction in nRMSE. Moreover, we introduce a novel weighting factor selection process to enhance generalizability and reconciliation accuracy. Lastly, we introduce power change as a prediction target, further improving the robustness and accuracy of CVR baseline estimation. Those considerations result in an additional reduction in nRMSE by 0.1-0.15\% in summer and 0.5-1.0\% in winter.

In this paper, we propose an \textbf{I}terative, \textbf{Bi}directional \textbf{G}radient \textbf{B}oosting \textbf{M}odel (IBi-GBM) for CVR baseline estimation. The \textbf{main contribution} of this paper is the development of IBi-GBM, a novel bidirectional estimation method featuring an iterative reconciliation process, which surpasses other existing algorithms with its exceptional accuracy, interpretability, and minimal historical data requirement. This method is designed to incrementally restore missing points, addressing two nearest points at a time, over an extended CVR period. This approach not only surpasses unidirectional predictions but also outperforms non-iterative methods commonly employed in existing algorithms. Because the CVR factor typically falls within a very narrow range (e.g., [0.3, 1.0]) \cite{wang2013review}, it has the potential to reside within the margin of error for regression-based models. To improve the estimation accuracy of IBi-GBM, we employ the following three techniques. \emph{Firstly}, to accurately capture the nonlinear behaviors of load, we employ GBM and LightGBM regressors, known for their ability to approximate any nonlinear functions \cite{yang2020lifespan}. \emph{Secondly}, to train the forward and backward estimations with high-quality inputs, we pre-screen non-CVR days based on the similarity of temperature and power profiles to the pre-CVR or post-CVR periods. This pre-screening process significantly improves prediction accuracy, enabling IBi-GBM to surpass deep-learning models even with a substantially smaller amount of training data. \emph{Lastly}, instead of predicting the total load ($P$), we estimate the load changes ($\Delta P$) between the previous point and the next point as the prediction target. This further enhances prediction accuracy by constraining forecasting errors within a specified range. 

The \textbf{second contribution} of this paper is that we provide a comparison between the bottom-up approach (utilizing aggregated smart meter data) and the top-down approach (employing SCADA data collected at substations) at various data resolutions to illustrate the influence of diverse data sources on the CVR baseline recovery process. This paper takes SCADA data metered at the substation as inputs and compares its performance with the approach using aggregated smart meter data as inputs so that the performance differences between the bottom-up and top-down approaches can be compared. Since most utilities have both types of data as inputs, using the two approaches for cross-validation can further decrease computing errors caused by inaccurate measurements and corrupted data, or act as a safeguard against intentional fake data injection attacks.

The importance of this research lies in its ability to empower utilities in selecting feeders for maximum load reduction through voltage reduction. This is achieved using an interpretable and computationally light model that demands a minimal amount of data. The rest of this paper is structured as follows: Section~\rom{2} provides an introduction to CVR and its performance evaluation, presenting the gradient boosting-based iterative bidirectional forecasting model and its application in analyzing the CVR effect. Simulation results are detailed in Section~\rom{3} and Section~\rom{4} serves as the conclusion of the paper.

\vspace{-0.15in}
\section{Methodology}
This section presents IBi-GB based baseline estimation algorithm and the performance evaluation criterion. 

\vspace{-0.15in}
\subsection{Description of CVR Event Data}
\begin{figure}[ht]
    \centerline{\includegraphics[width=\linewidth]{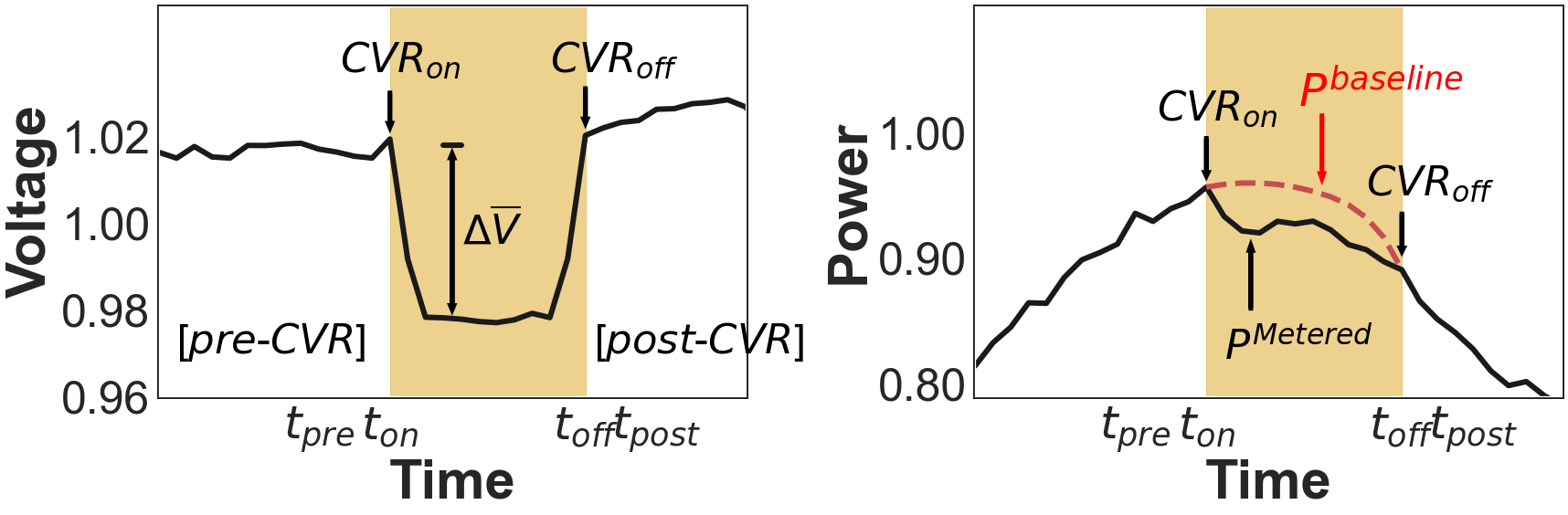}} 
    \vspace{-.2in}
    \subfloat[\label{1a}]{\hspace{.57\linewidth}}
    \subfloat[\label{1b}]{\hspace{.47\linewidth}} \vspace{-.05in}
    \caption{(a) Voltage profile and (b) load profiles in a CVR event.}
\label{fig1}
\vspace{-0.2in}
\end{figure}

As shown in Fig.~\ref{fig1}, during a CVR event, when the bus voltage is lowered by $\Delta \bar{V}\%$ at the distribution substation, we expect the real power consumption to change on average by $\Delta \bar{P}\%$. Thus, the CVR performance can be evaluated by the CVR factor, $CVR_{f}$ as
\begin{equation}
CVR_{f} = \frac{\Delta \bar{P}\%}{\Delta \bar{V}\%} = \frac{(\bar{P}^{baseline}-\bar{P}^{CVR})/\bar{P}^{baseline}}{(\bar{V}^{baseline}-\bar{V}^{CVR})/\bar{V}^{baseline}} \label{eq1}
\end{equation}
Note that in this paper, we focus on quantifying the real power reduction so the reactive power variations are ignored.  

Let $i$ ($i \in \{1,\cdots ,N^{nonCVR}\}$) and $j$ ($j \in \{1,\cdots,N^{CVR}\}$) be the indices of the non-CVR and CVR days, respectively. Thus, $P_{i}$ and $T_{i}$ are the power and temperature profiles for the $i^{th}$ non-CVR day, respectively; $P_{j}$ and $T_{j}$ are the power and temperature profiles for the $j^{th}$ CVR day, respectively. 

As shown in Fig.~\ref{fig1}, we divide a target CVR day into three periods: pre-CVR, CVR, and post-CVR. The CVR period is from $t_{on}$ to $t_{off}$ (the blue shaded area). Thus, $P^{BL}_{j}(t_{on}:t_{off})$ is the to-be-estimated CVR baseline in the $j^{th}$ CVR day. Select $N$ data samples immediately before and after the CVR event to be the pre-CVR and post-CVR periods. If the data resolution is $\Delta t$, the pre-CVR period is $t_{on} - N\times\Delta t$ to $t_{on}-\Delta t$, and the post-CVR period is from $t_{off}+\Delta t $ to $t_{off}+ N\times\Delta t$. 

\begin{figure*}[t!]
    \centerline{\includegraphics[width=\linewidth, height=.34\textheight]{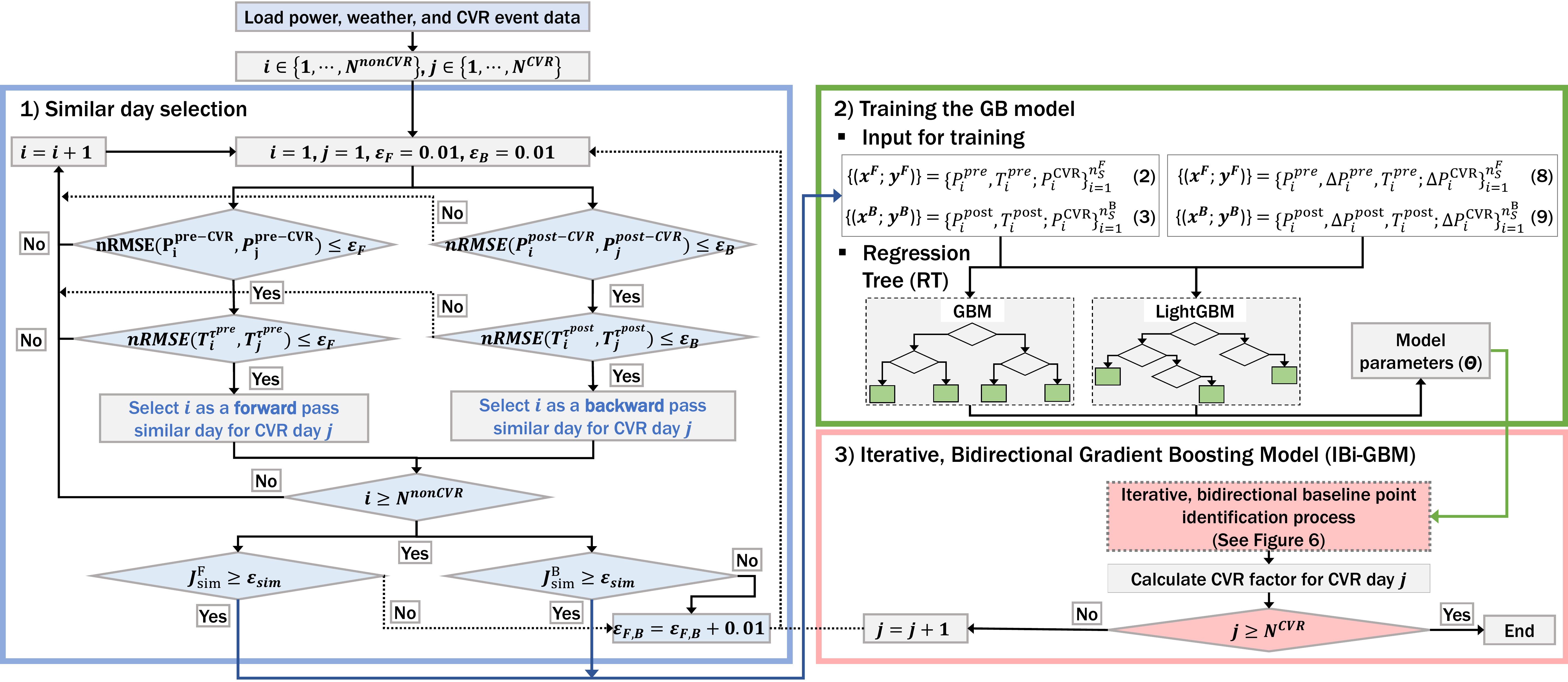}} \vspace{-0.1in}
    \caption{A three-stage framework proposed for CVR baseline estimation, involving: 1) selecting similar days, 2) training the GB model, and 3) implementing IBi-GBM.}
\label{fig2}
\vspace{-0.2in}
\end{figure*}

\vspace{-0.15in}
\subsection{Algorithm Overview}
As illustrated in Fig.~\ref{fig2}, the IBi-GB-based CVR baseline estimation methodology involves three essential processes: the selection of similar days, the training of the GB model, and the iterative generation of baseline data points. In the first process, pre-event and post-event temperature and power profiles of the targeted CVR day are used to select similar days from historical non-CVR days. Then, the pre-event and post-event similar days are used to train two GBMs: a forward-GBM and a backward-GBM. Lastly, the forward-GBM/backward-GBM generates the CVR baseline using pre-event/post-event load and temperature data as inputs, respectively. After each iteration, the two generated baselines are reconciled into one, where we keep only the first and the last points. Thus, two baseline points can be generated at each iteration. The iteration repeats until all CVR baseline points are generated. 

\begin{figure}[t!]
    \centerline{\includegraphics[width=\linewidth, height=.12\textheight]{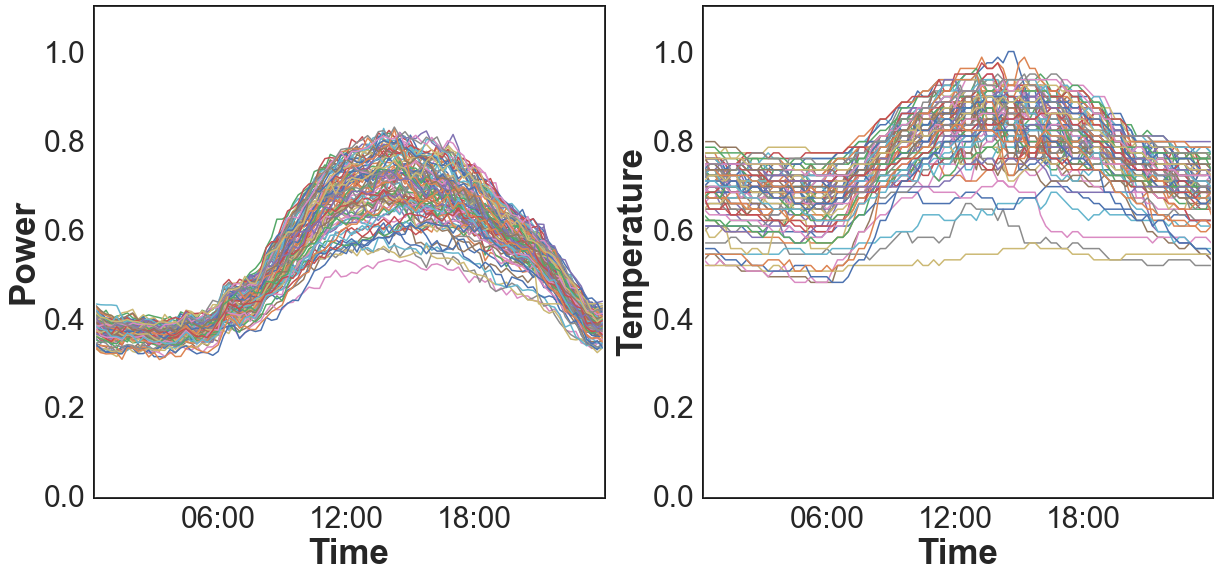}} \vspace{-.2in}
    \subfloat[\label{3a}]{\hspace{.57\linewidth}}
    \subfloat[\label{3b}]{\hspace{.43\linewidth}} \vspace{-.05in}
    \caption{(a) Normalized power profiles and (b) Normalized temperature profiles for the three summer months in 2020 of an actual feeder BR.}
\label{fig3}
\vspace{-0.2in}
\end{figure}

\begin{figure}[ht!]
    \centerline{\includegraphics[width=\linewidth, height=.19\textheight]{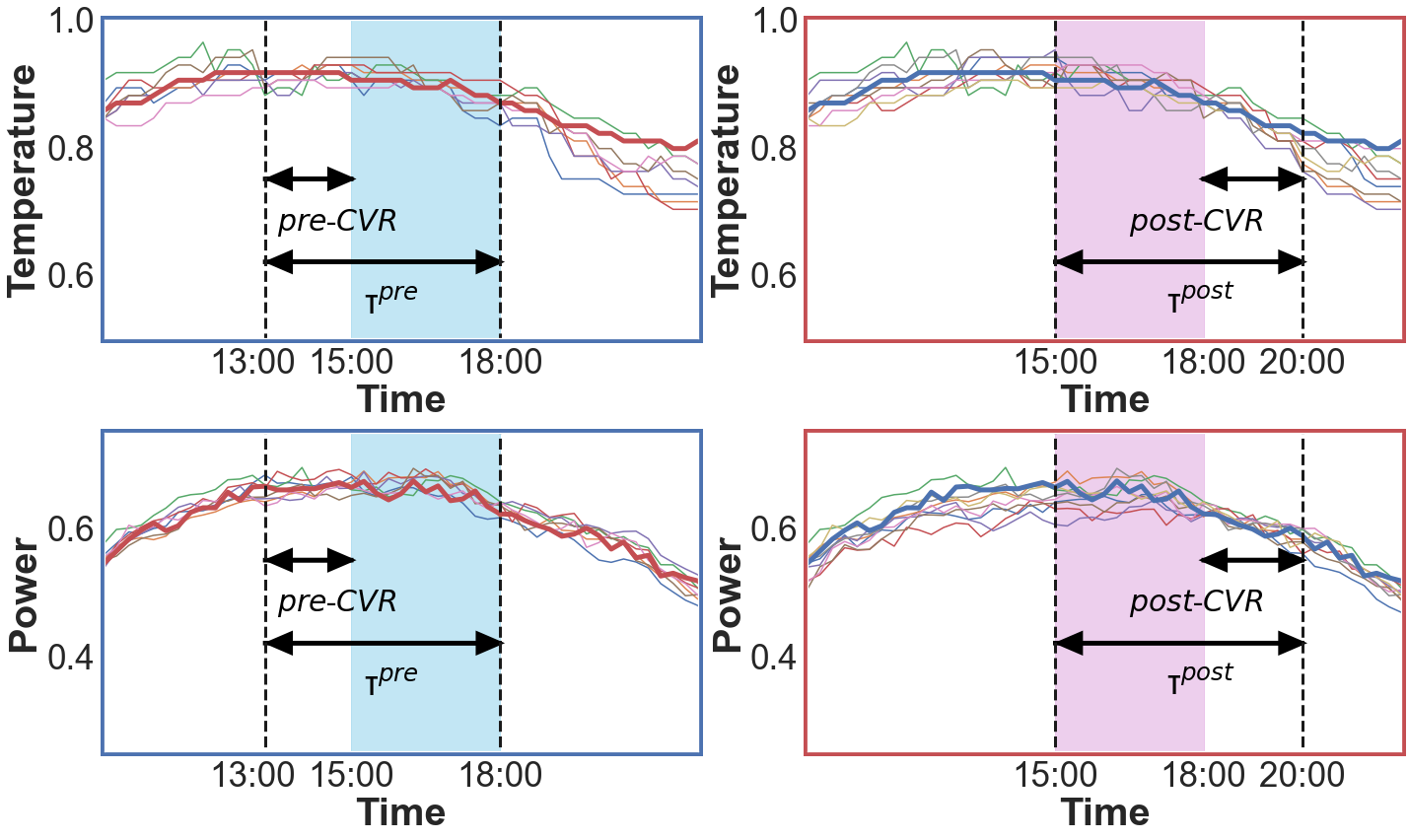}} \vspace{-.2in}
    \subfloat[\label{4a}]{\hspace{.57\linewidth}}
    \subfloat[\label{4b}]{\hspace{.43\linewidth}} \vspace{-.05in}
    \caption{Temperature and load profiles of the selected similar days for (a) the forward pass, and (b) the backward pass.}
\label{fig4}
\vspace{-0.3in}
\end{figure}

\vspace{-0.15in}
\subsection{Similar Day Selection}
To increase accuracy, training data should be selected from days with similar load changing patterns as those of the targeted CVR day. Conventional similar day selection \cite{wang2013analysis} is load based instead of weather based (i.e., outdoor temperature). This approach has a major drawback. If a CVR event lasts for more than one hour, the load can vary drastically due to sudden ambient temperature drops caused by cloud and precipitation. Selecting similar day by matching load profiles cannot account for such weather-dependent load variations. In addition, the load profile during a CVR event is unknown while the temperature and cloud information are known. Thus, by compensating for this factor and selecting days with matching load and weather profiles, the estimation accuracy can be greatly enhanced. 

As illustrated in Fig.~\ref{fig3} (a) and (b), the average Pearson Correlation Coefficient between power and temperature profiles of a real feeder is 0.73, showing very strong correlations in shapewise characteristics. Thus, in this paper, we propose a novel hybrid approach that utilizes both temperature and load-based criteria for similar day selection process to train the forward and backward gradient boosting-based models, respectively. Four sets of similar days are selected: the pre-event temperature similar days ($\Omega^{pre}_{T}$), the pre-event load similar days ($\Omega^{pre}_{P}$), the post-event temperature similar days ($\Omega^{post}_{T}$), and the post-event load similar days ($\Omega^{post}_{P}$). $\Omega^{pre}_{T}$ and $\Omega^{pre}_{P}$ are selected by matching temperature data in period $\tau^{pre}$ (from ($t_{on} - N\times\Delta t$) to $t_{off}$) and load data in period $pre$-$CVR$ (from ($t_{on} - N\times\Delta t$) to $t_{on}$), respectively. $\Omega^{post}_{T}$ and $\Omega^{post}_{P}$ are selected by matching temperature data in period $\tau^{post}$ (from $t_{on}$ to ($t_{off}+N\times\Delta t$)) and load data in period $post$-$CVR$ (from $t_{off}$ to ($t_{off}+N\times\Delta t$)), respectively.

As shown in Fig.~\ref{fig4}, four temperature and load segments in periods $\tau_{k}^{pre}$, $pre$-$CVR_{k}$, $\tau_{k}^{post}$, and $post$-$CVR_{k}$ in the $k^{th}$ targeted day are used to select similar days from non-CVR days ($N^{nonCVR}$) for training the forward and backward models, respectively. Compare the temperature and load segments in a non-CVR day in periods $\tau_{k}^{pre}$, $pre$-$CVR_{k}$, $\tau_{k}^{post}$, and $post$-$CVR_{k}$ with those in the targeted day by calculating nRMSE using \eqref{eq11}. If nRMSE for temperature and load both fall below a specified threshold (i.e., $\varepsilon_{F}$ and $\varepsilon_{B}$), a non-CVR day is selected as a similar day. In this paper, the $pre$-$CVR$ and $post$-$CVR$ windows are composed of $N \times \Delta t$ data points (e.g., 24 data points when using a 2-hour window with 5-min data), while both $\varepsilon_{F}$ and $\varepsilon_{B}$ are set to 1.0. To guarantee precise training of the proposed model, we introduced a minimum similar day constraint ($\varepsilon_{sim}$), set at 5, to avoid inaccuracies resulting from insufficient training data selection. The forward and backward training data for $n_{s}^{F}$ forward similar days and $n_{s}^{B}$ backward similar days can be represented as
\begin{IEEEeqnarray}{llC}
  \{(\bm{x^{F}}; \bm{y^{F}})\} \: &= \{(P_{i}^{pre}, T_{i}^{pre}; P_{i}^{CVR})\}_{i=1}^{n_{s}^{F}} \label{eq2} \\
  \{(\bm{x^{B}}; \bm{y^{B}})\} \: &= \{(P_{i}^{post}, T_{i}^{post}; P_{i}^{CVR})\}_{i=1}^{n_{s}^{B}} \label{eq3}
\vspace{-0.1in}
\end{IEEEeqnarray}

\subsection{Gradient Boosting Models}
\begin{figure}[ht!]
    \centerline{\includegraphics[width=0.95\linewidth, height=0.11\textheight] {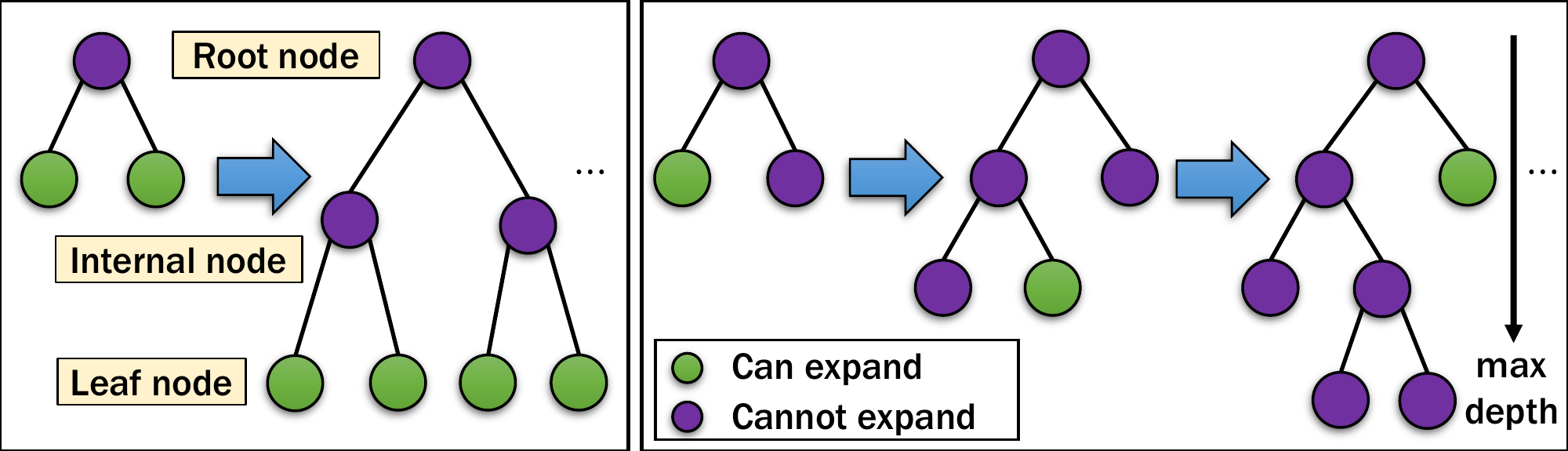}} \vspace{-.2in}
    \subfloat[\label{5a}]{\hspace{.46\linewidth}}
    \subfloat[\label{5b}]{\hspace{.46\linewidth}} \vspace{-0.05in}
    \caption{In gradient boosting algorithms, (a) refers to GBM, characterized by level-wise tree growth, and (b) refers to LightGBM, distinguished by leaf-wise tree growth.}
\label{fig5}
\vspace{-0.2in}
\end{figure}

In this paper, we compare two types of gradient boosting-based regression models: GBM and LightGBM for four data resoultions (i.e., 5-, 15-, 30-, and 60-minute). Both GBM and LightGBM are decision tree-based methods capable of approximating nonlinear functions. As shown in Fig.~\ref{fig5}, GBM splits the tree level-wise for the best fit, while the LightGBM algorithm splits the tree leaf-wise. GBM is a highly efficient method for constructing predictive models for low-dimensional data \cite{natekin2013gradient} while LightGBM runs faster and is a more distributed approach when handling high-dimensional data \cite{ke2017lightgbm}. For detailed descriptions of the two GBM algorithms, please refer to \cite{natekin2013gradient, wang2020ss}.
 
\vspace{-0.1in}
\subsection{IBi-GB-based Baseline Identification Algorithm} 
For a CVR event lasting for hours, the baseline prediction accuracy drops sharply towards the end when using unidirectional method. Thus, we develop an iterative, bidirectional prediction process to replace the conventional unidirectional prediction approach. 

\begin{figure}[t!]
    \centerline{\includegraphics[width=\linewidth, height=.25\textheight]{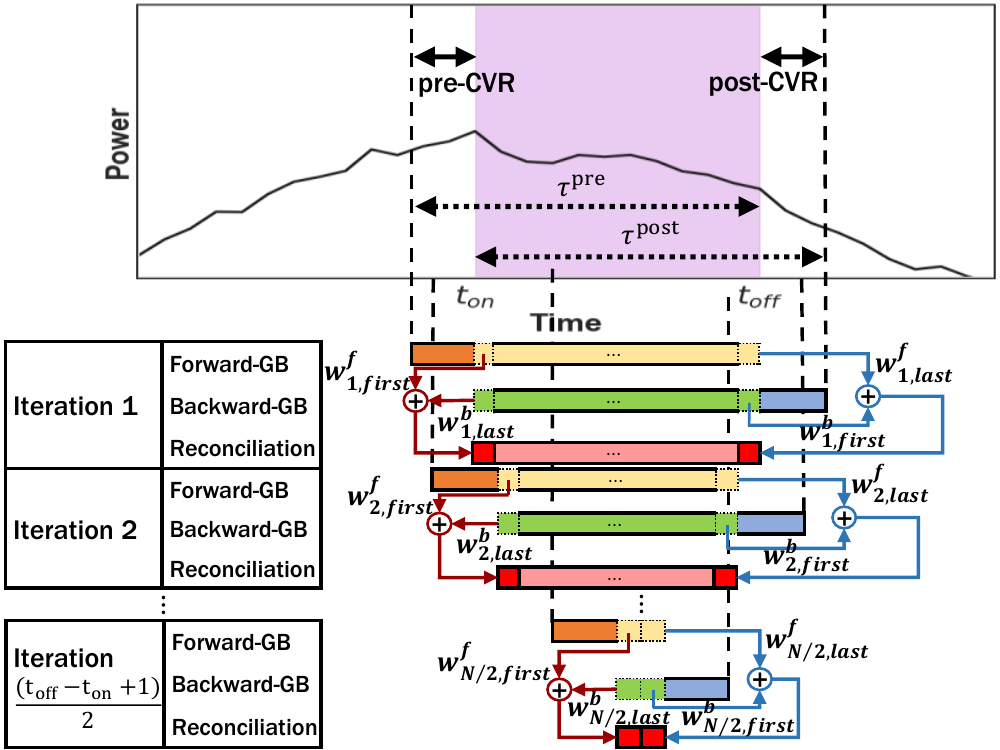}} \vspace{-0.1in}
    \caption{An illustration of the iterative, bidirectional gradient boosting model (IBi-GBM) for the process of CVR baseline estimation.}
\label{fig6}
\vspace{-0.3in}
\end{figure}

\subsubsection{Reconciliation Weighting factors derivation}
To reconcile the outcomes of forward and backward forecasting results, a multiple linear regression process is used to determine two sets of weighting factors: one for the forward direction ($w^{f}$) and another for the backward direction ($w^{b}$). Select $K$ virtual-CVR days that are similar to the given CVR day, and let $k$ represent the $k^{th}$ day. The process for computing the $i^{th}$ reconciliation weighing factors for the $k^{th}$ virtual-CVR day can be formulated as follows: 
\vspace{-0.1in}
\begin{IEEEeqnarray}{llC} \label{eq4}  
     P^{GT}_{k,1}  &= \beta_{0} + \beta_{1,1}\hat{P}^{f}_{k,1} + \beta_{2,N}\hat{P}^{b}_{k,N} \nonumber \vspace{-0.1in} \\
     &\vdots \nonumber \\ 
      P^{GT}_{k,i}  &= \beta_{0} + \beta_{1,i}\hat{P}^{f}_{k,i} + \beta_{2,N-i+1}\hat{P}^{B}_{k,N-i+1} \nonumber   \vspace{-0.1in} \\
     &\vdots \nonumber \\
      P^{GT}_{k,N}  &= \beta_{0} + \beta_{1,N}\hat{P}^{f}_{k,N} + \beta_{2,1}\hat{P}^{B}_{k,1}  \vspace{-0.1in}
\end{IEEEeqnarray}
where $N$ is the number of data points in the CVR period, $P_k^{GT}$, $\hat{P}_{k}^{f}$, and $ \hat{P}_{k}^{b}$ are $1\times N$ matrices representing the ground truth, the forward and backward forecasted CVR baseline, respectively; $\beta_{0}$, $\beta_{1}$, and $\beta_{2}$ are $1\times N$ matrices containing the regression coefficients for each data point. 

As the forward and backward predictions are optimized to match the ground truth, we set $\beta_{0}$ values to be zero. Thus, $\beta_{1}$ and $\beta_{2}$ can be viewed as the forward and backward weighting factors for merging the two predictions into the ground truth. 

The process for computing the reconciliation weighting factors optimized by ground truth profiles from all $K$ days can be expressed as 
\vspace{-0.1in}
\begin{IEEEeqnarray}{llC} 
    (\bm{Y_k})^T = (\beta_{1})^T\cdot(\bm{X_k^{F}})^T + (\beta_{2})^T\cdot(\bm{X_k^{B}})^T, \nonumber \\
    \bm{Y} =
    \begin{bmatrix}
    P^{GT}_{1} \vspace{-0.1in} \\
    \vdots \\ P^{GT}_{k} \vspace{-0.1in} \\
    \vdots \\
    P^{GT}_{K}
   \end{bmatrix}, 
    \bm{X^F} =
    \begin{bmatrix}
    \hat{P}^{f}_{1} \vspace{-0.1in} \\
    \vdots \\
    \hat{P}^{f}_{k} \vspace{-0.1in} \\
    \vdots \\
    \hat{P}^{f}_{K}
    \end{bmatrix}, 
    \bm{X^B} =
    \begin{bmatrix}
    \hat{P}^{b}_{1} \vspace{-0.1in} \\
    \vdots \\
     \hat{P}^{b}_{k} \vspace{-0.1in} \\
    \vdots \\   
    \hat{P}^{b}_{K}
    \end{bmatrix}   \vspace{-0.1in} \label{eq5}
\end{IEEEeqnarray}
where $\bm{Y}$, $\bm{X^F}$, and $\bm{X^B}$ contains the ground truth profiles, the forward and backward predictions of the CVR baselines for all $K$ virtual CVR days. 

Note that this procedure is applicable for computing the sets of weighting factors in each of the $\frac{t_{off}-t_{on}+1}{2}$ iterations. Since the length of the forecasting data ($N$) is known for each iteration, the weighting factors can be calculated offline. As demonstrated in the Results section, the values of weighting factors do not vary significantly across various substations for a given length of the forecasting data. Consequently, once the weighting factors are determined, they can be applied across substations.

\begin{figure}[b!]
\vspace{-.2in}
 \centerline{\includegraphics[width=\linewidth, height=.12\textheight]{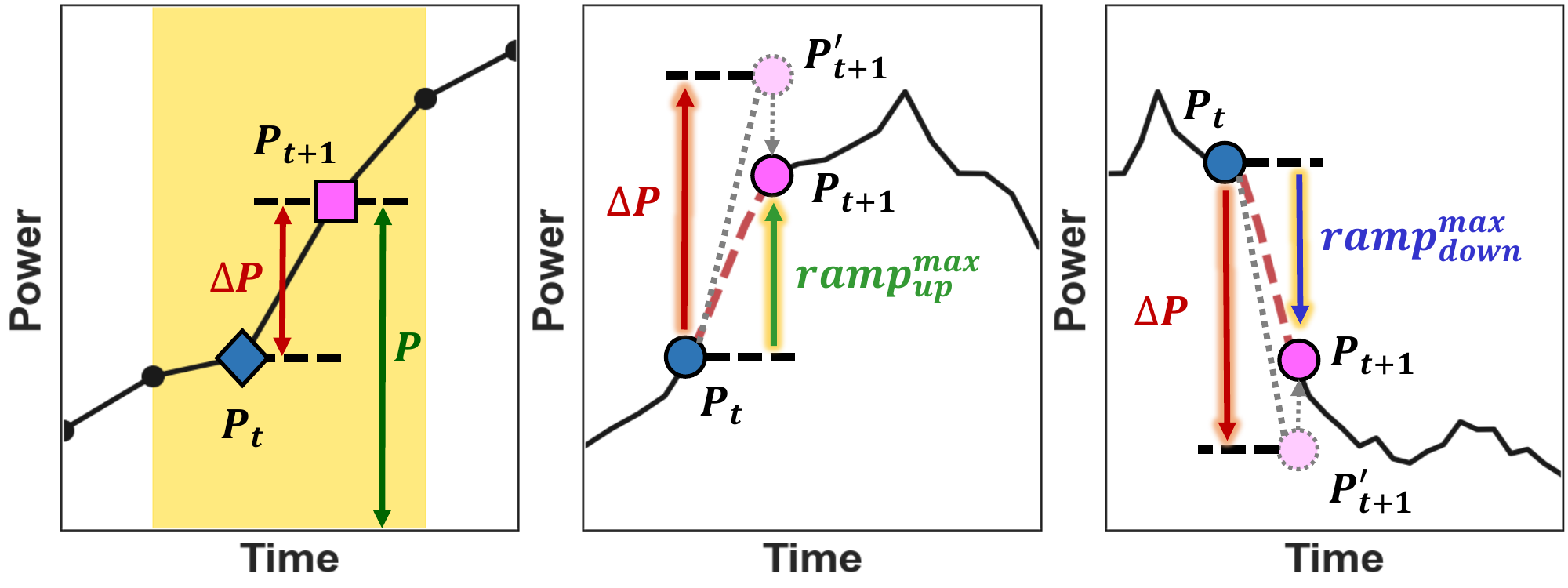}} 
 \vspace{-.225in}
    \subfloat[\label{7a}]{\hspace{.38\linewidth}}
    \subfloat[\label{7b}]{\hspace{.28\linewidth}} 
    \subfloat[\label{7c}]{\hspace{.38\linewidth}} \vspace{-.075in}
    \caption{An illustration of the error bounding mechanism. (a) Target 1: predicting load ($P$) and target 2: predicting load change ($\Delta P$), (b) Bounding error by the maximum ramp-up constraint ($ramp^{max}_{up}$), and (c) Bounding error by the maximum ramp-down constraint ($ramp^{max}_{down}$).}
\label{fig7}
\end{figure}

\begin{table*}[ht!]
    \begin{center}
    \caption{Description of the Testing Data.}
    \vspace{-.2in}
    \label{tab2}
    \resizebox{1.0\linewidth}{!}{%
    \begin{tabularx}{1.04\linewidth}{ccccccccccccc}    
    \toprule
    \multirow{2}{4em}{\parbox{1\linewidth}{\centering \textbf{Data} \\ \textbf{Source}}}
    &\multirow{2}{*}{\vspace{-0.1cm} \centering \textbf{Utility}} 
    &\multirow{2}{*}{\vspace{-0.1cm} \centering \textbf{SS}} 
    &\multirow{2}{*}{\vspace{-0.1cm} \centering \textbf{Feeder}} 
    &\multirow{2}{4em}{\parbox{1\linewidth}{\vspace{0.1cm} \centering \textbf{Data} \\ \textbf{Rez.}}} 
    &\multirow{2}{4em}{\parbox{1\linewidth}{\vspace{0.1cm} \centering \textbf{Data} \\ \textbf{Length}}}
    &\multicolumn{2}{c}{\textbf{CVR day}} & &\multicolumn{2}{c}{\textbf{Virtual CVR day}}
    &\multirow{2}{4em}{\parbox{1\linewidth}{\vspace{0.1cm} \centering \textbf{CVR} \\ \textbf{duration}}} 
    &\multirow{2}{5em}{\parbox{1\linewidth}{\vspace{0.1cm} \centering \textbf{non-CVR} \\ \textbf{day}}} \\
    \cmidrule{7-8} \cmidrule{10-11}
    & & & & & &{\textbf{Winter}} &{\textbf{Summer}} & &{\textbf{Winter}} &{\textbf{Summer}} \\
    \midrule \midrule
    1 &\multirow{3}{*}{\vspace{-0.3cm} \centering A} &WC &BR &\multirow{3}{*}{\vspace{-0.3cm} \centering \makecell{15-min \\ smart meter \\ data}} &\multirow{3}{*}{\vspace{-0.3cm} \centering \makecell{731 \\ days}}  &\multirow{3}{*}{\vspace{-0.3cm} \centering -} &\multirow{3}{*}{\vspace{-0.3cm} \centering 24} & &\multirow{3}{*}{\vspace{-0.3cm} \centering -}   &\multirow{3}{*}{\vspace{-0.3cm} \centering 103} &\multirow{3}{*}{\vspace{-0.3cm} \centering \makecell{Fixed \\ 180 min}} &677  \\
    \cmidrule{3-4} \cmidrule{13-13}
    2 & &\multirow{2}{*}{\centering SF} &DF & &  & & & &   & & &679  \\
    \cmidrule{4-4} \cmidrule{13-13}
    3 & & &SL & &  & & & &   & & &679  \\
    \midrule
    4 &\multirow{3}{*}{\vspace{-1.0cm} \centering B} &CL &- &\multirow{3}{*}{\vspace{-1.0cm} \centering \makecell{5-min \\ SCADA\\ data}} &\multirow{3}{*}{\vspace{-1.0cm} \centering \makecell{1339\\ days}} &\multirow{3}{*}{\vspace{-1.0cm} \centering 7} &\multirow{3}{*}{\vspace{-1.0cm} \centering 48} & &\multirow{3}{*}{\vspace{-1.0cm} \centering 137} &\multirow{3}{*}{\vspace{-1.0cm} \centering 140} &\multirow{3}{*}{\vspace{-1.0cm} \centering \makecell{ Min. \\ 60 min \\ \\ Max. \\ 180 min}} &645 \\
    \cmidrule{3-4} \cmidrule{13-13}
    5 & &RF &\makecell{913, 914, 915, \\ 916, 917, 918} &  & & & & & & & &661 \\
    \cmidrule{3-4} \cmidrule{13-13}
    6 & &RR &\makecell{760, 761, 762, \\ 763, 764, 765} &  & & & & & & & &730 \\
    \bottomrule
    \end{tabularx}}
    \end{center}
\vspace{-.35in}
\end{table*}

\subsubsection{Forecast Reconciliation}
As shown in Fig.~\ref{fig6}, at the beginning of the $i^{th}$ iteration, the forward-pass input data in the orange segment, which includes the pre-CVR data in a 4-hour window, is used to generate the first set of CVR baseline data points in yellow ($\hat{P}^{f}$). Then, the backward-pass input data set in the blue segment, which includes the post-CVR data in a 4-hour window, is used to generate the second set of CVR baseline data points in green ($\hat{P}^{b}$). 

Next, take the first and the last data points from the forward and the backward forecasting results and reconcile them by
\vspace{-0.05in}
\begin{IEEEeqnarray}{llC}
   \hat{P}_{i}^{R} &= w_{i,first}^{f} \times \hat{P}_{first}^{f} + w_{i,last}^{b} \times \hat{P}_{last}^{b} \label{eq6} \\    
   \hat{P}_{N-i+1}^{R} &= w_{i,first}^{b} \times \hat{P}_{first}^{b} + w_{i,{last}}^{f} \times \hat{P}_{last}^{f} \label{eq7}
\vspace{-0.1in}
\end{IEEEeqnarray}
where $\hat{P}_{first}^{f}$ and $\hat{P}_{last}^{f}$ are the first and last forecasted values in the forward forecasting results, $\hat{P}_{first}^{b}$ and $\hat{P}_{last}^{b}$ are the first and last forecasted values in the backward forecasting results, and $\hat{P}_{i}^{R}$ and $\hat{P}_{N-i+1}^{R}$ are the reconciled values for the $i^{th}$ iteration. $N$ is the total number of points in the baseline.

Please note that $w_{i,first}^{f}$, $w_{i,last}^{f}$, $w_{i,first}^{b}$, and $w_{i,last}^{b}$ represent the weighting factors corresponding to the first and last forecasted values in the forward and backward forecasting results, respectively. The weighting factors are derived from the first and last values of $\beta_{1}$ for forward weightings and $\beta_{2}$ for backward weightings, as calculated by \eqref{eq5}. The two reconciled data points are then incorporated into the input data for the following iteration, as shown in Fig.~\ref{fig6}. The progression to the next iteration is facilitated by advancing both the forward and backward data windows by a single data point. 

\subsubsection{Prediction Targets}
As shown in Figs.~\ref{fig7}(a), to further enhance the robustness and accuracy of CVR baseline estimation, instead of predicting load ($P$), we can predict load change ($\Delta P$) by replacing the forward and backward training data in \eqref{eq2} and \eqref{eq3} by

\vspace{-0.2in}
\begin{IEEEeqnarray}{llC}
  \{(\bm{x^{F}}; \bm{y^{F}})\} \: &= \{(P_{i}^{pre}, \Delta P_{i}^{pre}, T_{i}^{pre}; \Delta P_{i}^{CVR})\}_{i=1}^{n_{s}^{F}} \label{eq8} \\
  \{(\bm{x^{B}}; \bm{y^{B}})\} \: &= \{(P_{i}^{post}, \Delta P_{i}^{post}, T_{i}^{post}; \Delta P_{i}^{CVR})\}_{i=1}^{n_{s}^{B}} \label{eq9}
\end{IEEEeqnarray}

Setting $\Delta P$ as the prediction target can effectively limit the prediction error. At the substation and feeder levels, sudden changes in load are rare, except for instances of failures or the addition/loss of substantial loads. Thus, by extracting the maximum upward and downward changes (i.e., $ramp^{max}_{up}$ and $ramp^{max}_{down}$) in $\Delta P$ from historical data, forecasts outside this range can be adjusted to fall within the specified bounds, as illustrated in Figs.~\ref{fig7}(b) and (c).

\vspace{-.15in}
\subsection{CVR Performance Evaluation Criterion}
The estimation performance can be evaluated by mean absolute percentage error (MAPE), normalized root mean squared error (nRMSE), energy error ($\epsilon_{e}$), and mean percentage error (MPE). The metrics are defined as
\begin{IEEEeqnarray}{llC}
  MAPE &= \frac{1}{N} \sum_{n} \sum_{t} \left| \frac{y_{t}^{n}-\hat{y}_{t}^{n}}{y_{t}^{n}} \right| \label{eq10} \\
  nRMSE\ &= \frac{1}{N} \sum_{n} \sqrt {\sum_{t} \frac{(y_{t}^{n}-\hat{y}_{t}^{n})^2}{T}} \bigg/ \left( \frac{\sum_{t} y_{t}^{n}}{T} \right) \label{eq11} \\
  \epsilon_{e} &= \frac{1}{N} \sum_{n} \frac{\sum_{t} \left| y_{t}^{n} - \hat{y}_{t}^{n} \right|}{\sum_{t} y_{t}^{n}} \label{eq12} \\
  MPE &= \frac{100\%}{N} \sum_{n} \left( \frac{1}{T} \sum_{t} \frac{y_{t}^{n} - \hat{y}_{t}^{n}}{y_{t}^{n}} \right)  \label{eq13} 
\end{IEEEeqnarray}
where $N$ is the total number of test days, $y_{t}^{n}$ is the actual value at time step $t$ on test day $n$, $\hat{y}_{t}^{n}$ is the predicted value at time step $t$ on test day $n$.

\vspace{-.15in}
\section{Performance Evaluations}
In this section, we compare the performance of IBi-GBM with the unidirectional model (i.e., MSVR \cite{wang2013analysis}) and a GAN-based MSD restoration model (i.e., Load-PIN \cite{li2023load}). We also present the impact of input data resolution on the IBi-GBM prediction accuracy considering variable CVR duration. The CVR efficacy for a few actual feeders in a distribution substation is also presented.

\begin{table}[t!]
\vspace{-.05in}
    \begin{center}
    \caption{Hyper-parameters of Gradient Boosting Models.}
    \vspace{-.1in}
    \label{tab3}
    \resizebox{0.9\linewidth}{!}{%
    \begin{tabularx}{1.05\linewidth}{ccccccc}    
    \toprule
    &\multirow{2}{7.2em}{\parbox{\linewidth}{\vspace{0.1cm} \centering \textbf{Hyper} \\ \textbf{parameters ($\bm{\theta_{h}}$)}}}
    &\multirow{2}{*}{\centering \textbf{Utility}} 
    &\multicolumn{4}{c}{\textbf{Data resolution}} \\   
    \cmidrule{4-7}
    & & &{\textbf{5-min}} &{\textbf{15-min}} &{\textbf{30-min}} &{\textbf{60-min}} \\
    \midrule \midrule
    \multirow{4}{*}{\rotatebox[origin=c]{90}{\hspace{-0.1cm} IBi-GBM}} & 
    \multirow{2}{7em}{\parbox{\linewidth}{\centering \textbf{learning rate} \\ \textbf{($\alpha$)}}} 
      &A &-      &0.1    &0.1    &0.075 \\
    & &B &0.1    &0.075  &0.1    &0.1   \\
    \cmidrule{2-7}
    & \multirow{2}{7em}{\parbox{\linewidth}{\centering \textbf{n estimators} \\ \textbf{($N_{E}$)}}} 
      &A &-      &100    &75    &50   \\
    & &B &200    &150    &100   &50   \\
    \midrule
    \multirow{6}{*}{\rotatebox[origin=c]{90}{\hspace{-0.3cm} IBi-LightGBM}} & 
    \multirow{2}{7em}{\parbox{\linewidth}{\centering \textbf{learning rate} \\ \textbf{($\alpha$)}}}
      &A &-      &0.1    &0.1   &0.075 \\
    & &B &0.1    &0.075  &0.1   &0.05  \\
    \cmidrule{2-7}    
    & \multirow{2}{7em}{\parbox{\linewidth}{\centering \textbf{n estimators} \\ \textbf{($N_{E}$)}}}
      &A &-      &100   &75     &50    \\
    & &B &200    &150   &100    &50    \\
    \cmidrule{2-7}
    & \multirow{2}{7em}{\parbox{\linewidth}{\centering \textbf{max depth} \\ \textbf{($d_{max}$)}}}
      &A &-     &5      &5      &5     \\
    & &B &7     &7      &7      &7     \\
    \bottomrule
    \end{tabularx}}
    \end{center}
\vspace{-.35in}
\end{table}

\vspace{-.2in}
\subsection{Data Preparation}
There are two data sets used to develop and verify the performance of two iterative bidirectional-GB based CVR baseline estimation algorithms. The first data set contains smart meter data collected by utility A in North Carolina in 2019 and 2020. By aggregating 15-min smart meter data on the same feeder together, we obtain the total load profile for three feeders. Thus, the CVR load reduction computed in this data is the net load reduction on the customer side so that transformer losses and line losses are not included. The second data set contains four years of 5-min SCADA data from 2019 to 2022 for twelve feeders at three substations (i.e., CL, RF, and RR) collected by utility B in North Carolina. A detailed description of the data is provided in Table~\ref{tab2}. Thus, in total, we compare the algorithm performance on three substations and fifteen feeders using both substation, feeder head, and smart meter data with data resolution ranging from 5-min to 60-min. 

To quantify the algorithm's performance, we select \emph{virtual CVR days}, which refer to days excluding CVR days and holidays during the three summer and winter months of all years, from non-CVR days. During these selected days, we assume that CVR is implemented for 3 hours (from 15:00 to 18:00) in summer and 1.5 hours (from 07:00 to 08:30) in winter. Thus, on a virtual CVR day, the load profile during the targeted CVR period (i.e., the CVR baseline) is known. Once the CVR baseline is estimated using the proposed algorithm, it can be compared with the ground truth load profile ($P^{GT}$) to compute prediction accuracy.

Unlike the model parameters $\bm{\Theta}$ determined in the training process, the hyperparameters ($\bm{\theta_{h}}$) should be specified before training. Since a regression tree (RT) model has fewer parameters, the Grid Search algorithm is used for their tuning. We consider three hyperparameters: (a) $\alpha$ is learning rate (i.e., shrinking the contribution of individual RTs), (b) $N_{E}$ is the number of estimator, and (c) $d_{max}$ is the maximum depth of RT (i.e., number of edges with the farthest distance between the root split node and the leaf node, e.g., $d_{max}$ = 3 in the Fig.~\ref{fig5}(b). The $\bm{\theta_{h}}$ of each model are summarized in Table~\ref{tab3}. The remaining parameters adhere to the default settings defined in Python Scikit Learn \cite{pedregosa2011scikit}.

\begin{figure}[b!]
    \vspace{-.2in}
    \subfloat[]{%
        \centerline{\includegraphics[width=0.98\linewidth, height=0.1\textheight]{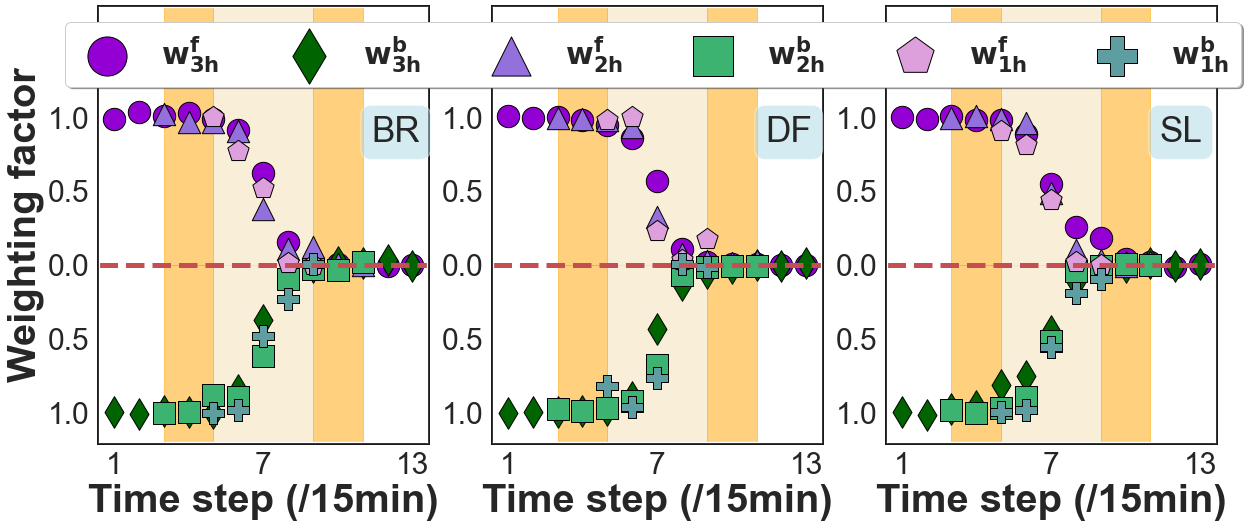}}%
        \label{8a}%
    }\hfill \vspace{-.1in} \\
    \subfloat[]{%
        \centerline{\includegraphics[width=0.98\linewidth, height=0.1\textheight]{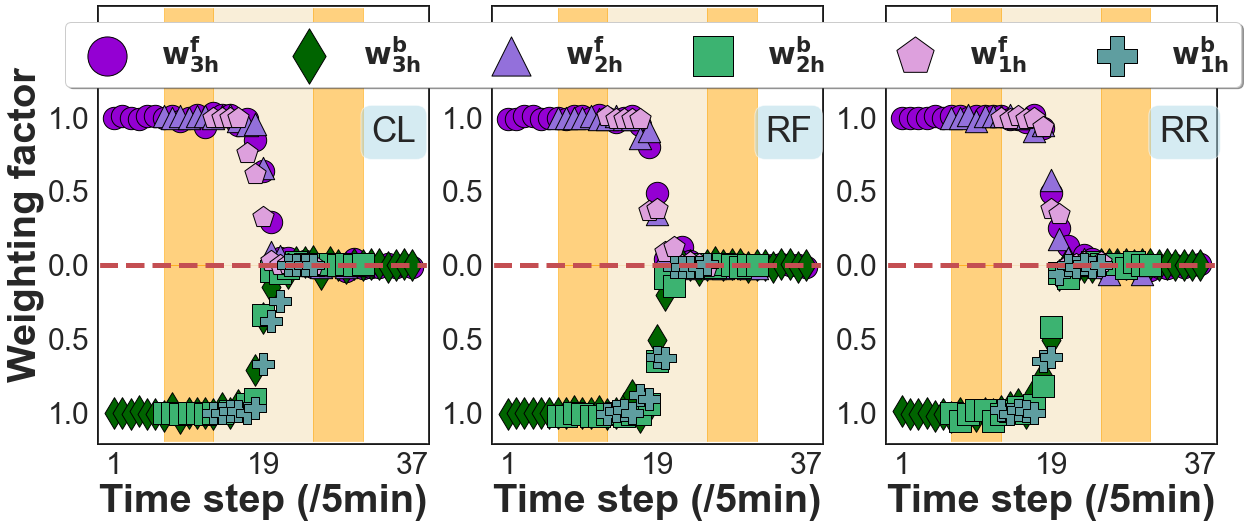}}%
        \label{8b}%
    }\hfill \vspace{-.1in} \\
    \subfloat[]{%
        \centerline{\includegraphics[width=0.98\linewidth, height=0.1\textheight]{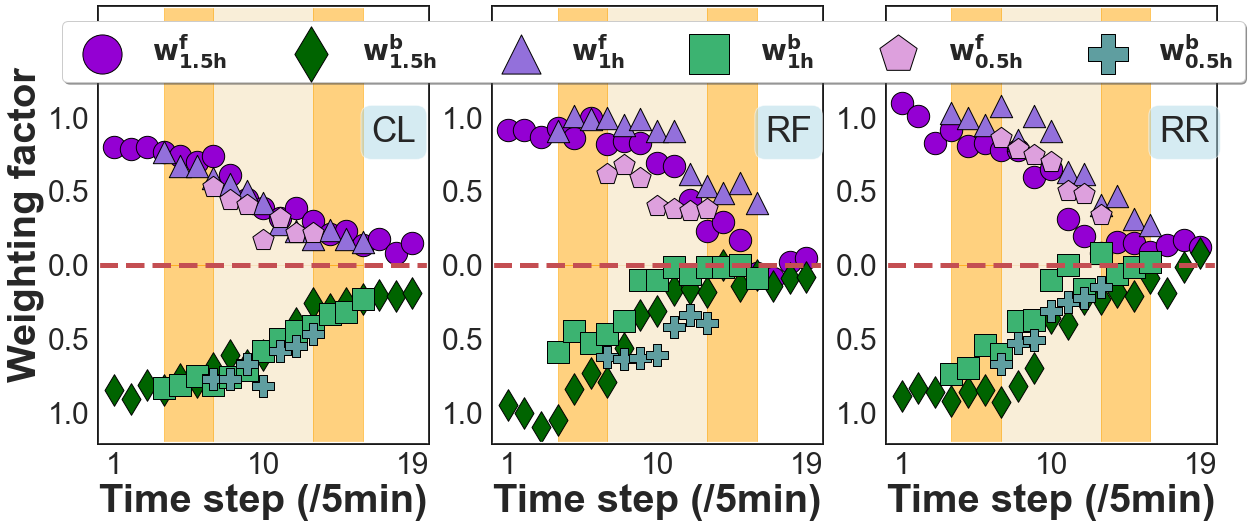}}%
        \label{8c}%
    }
    \vspace{-0.05in}    
    \caption{Reconciliation weighting factors of the virtual CVR days for (a) three feeders with 15-min smart meter data, (b) three substations with 5-min SCADA data in summer, and (c) three substations with 5-min SCADA data in winter.}
    \label{fig8}
\vspace{-.1in}
\end{figure}

\vspace{-0.2in}
\subsection{Performance Evaluation on the \textbf{Virtual} CVR day}
To evaluate the efficacy of the proposed algorithm, we select 103 virtual CVR days from Utility A's dataset, along with 137 virtual CVR days in winter and 140 virtual CVR days in summer from Utility B's dataset. For benchmarking the IBi-GBM and IBi-LightGBM models, we evaluate their performance against regression-based and deep learning-based models, specifically MSVR (Multistage SVR) \cite{wang2013analysis} and Load-PIN (Load Profile Inpainting) \cite{li2023load}, respectively.

\begin{figure}[t]
    \subfloat[]{%
        \centerline{\includegraphics[width=0.98\linewidth, height=0.1\textheight]{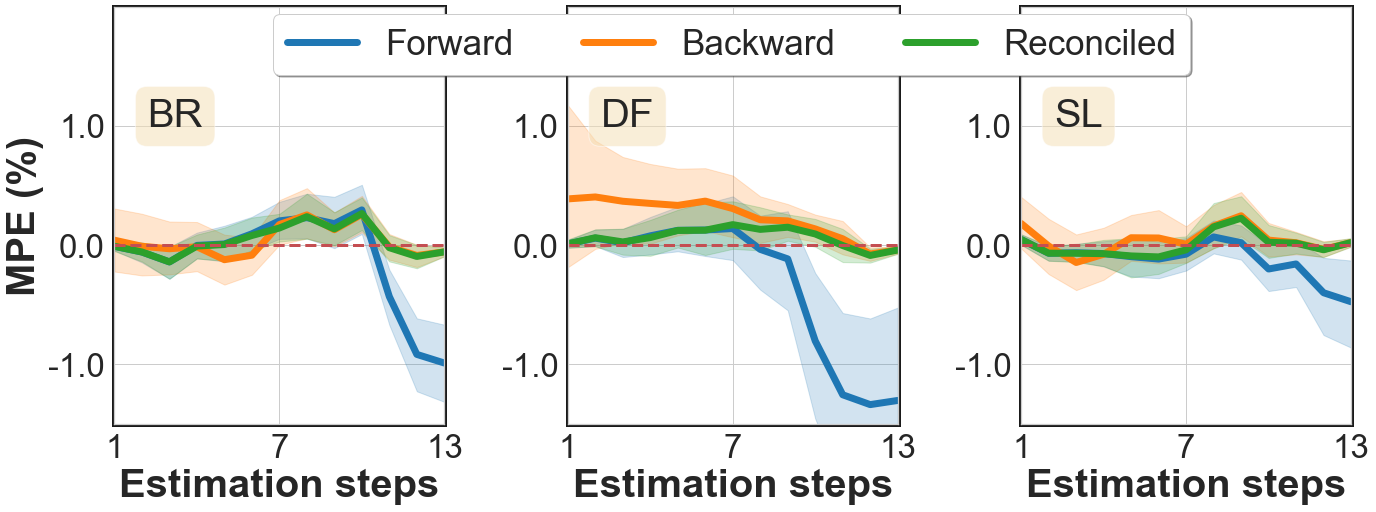}}%
        \label{9a}%
    }\hfill \vspace{-.1in} \\
    \subfloat[]{%
        \centerline{\includegraphics[width=0.98\linewidth, height=0.1\textheight]{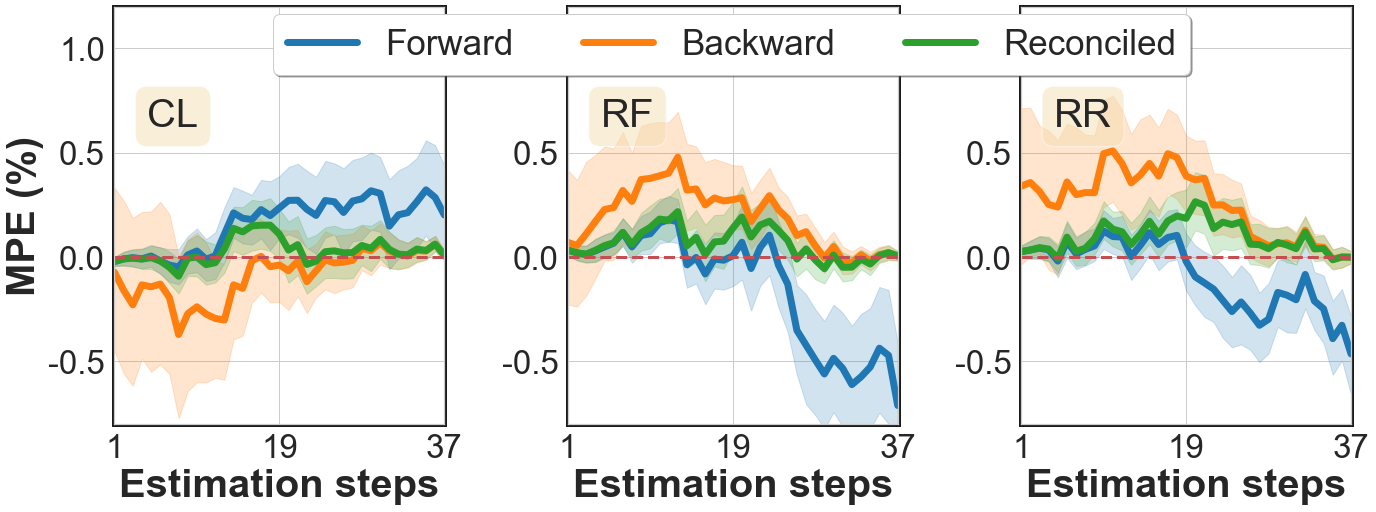}}%
        \label{9b}%
    }\hfill \vspace{-.1in} \\
    \subfloat[]{%
        \centerline{\includegraphics[width=0.98\linewidth, height=0.1\textheight]{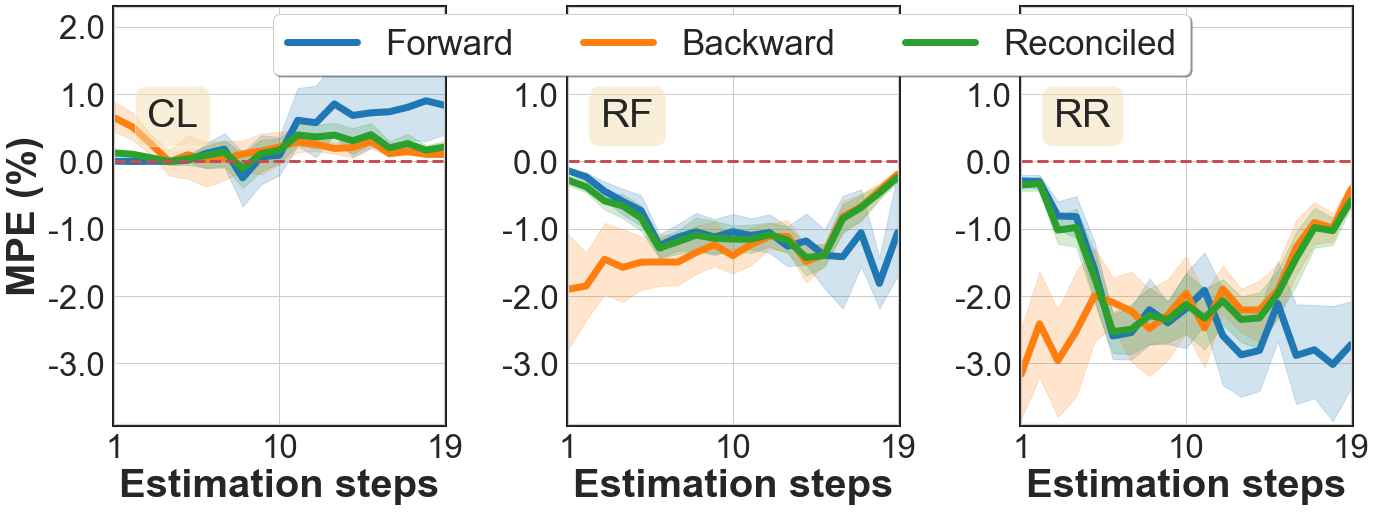}}%
        \label{9c}%
    }
    \vspace{-0.05in}    
    \caption{Estimation accuracy and 95\% confidence intervals of the forward, backward, and reconciled approaches for (a) three feeders with 15-min smart meter data, (b) three substations with 5-min SCADA data in summer, and (c) three substations with 5-min SCADA data in winter.} 
    \label{fig9}
\vspace{-.2in}
\end{figure}

\begin{figure}[b]
\vspace{-0.2in}
\centerline{\includegraphics[width=\linewidth, height=0.1\textheight]{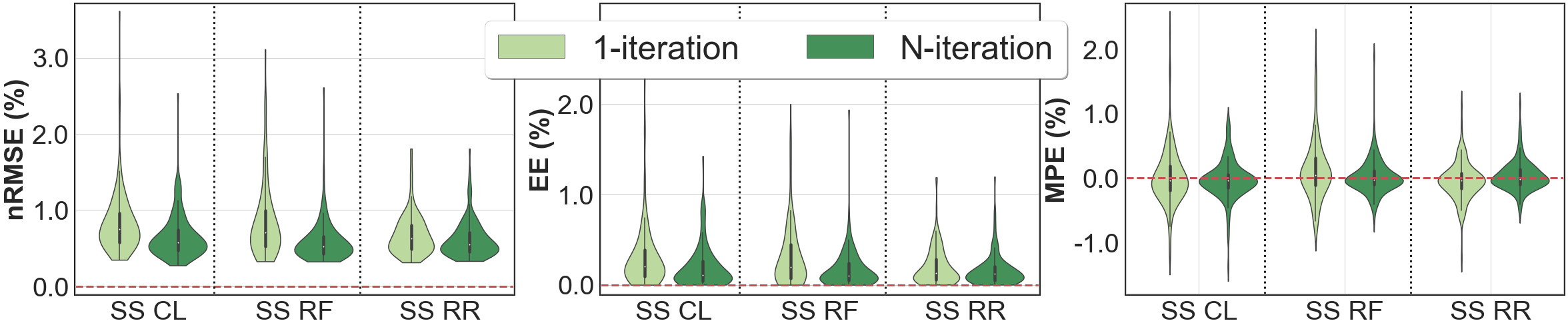}} \vspace{-.2in}
  \subfloat[\label{10a}]{\hspace{.38\linewidth}}
  \subfloat[\label{10b}]{\hspace{.29\linewidth}} 
  \subfloat[\label{10c}]{\hspace{.38\linewidth}} 
\vspace{-.05in}
\caption{Violin plots for comparing the prediction accuracy of one-time and iterative reconciliations for three substations with 5-min SCADA data during summer months: (a) nRMSE, (b) EE, and (c) MPE.}
\label{fig10}
\vspace{-.1in}
\end{figure}

\begin{table*}[ht]
\begin{center}
    \caption{Average Estimation Errors of Virtual-CVR Days at Feeder Level} \vspace{-0.1in}
    \label{tab4}
    \centerline{\includegraphics[width=\linewidth, height=0.16\textheight]{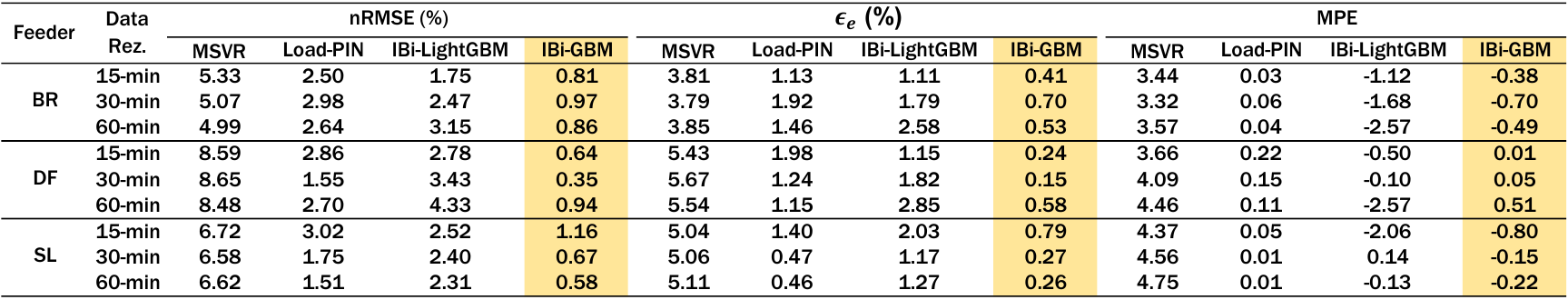}}    
\end{center}
\vspace{-0.4in}
\end{table*}

\begin{table*}[ht]
\begin{center}
    \caption{Average Estimation Errors of Virtual-CVR Days at Substation Level} \vspace{-0.1in}
    \label{tab5}
    \centerline{\includegraphics[width=\linewidth, height=0.35\textheight]{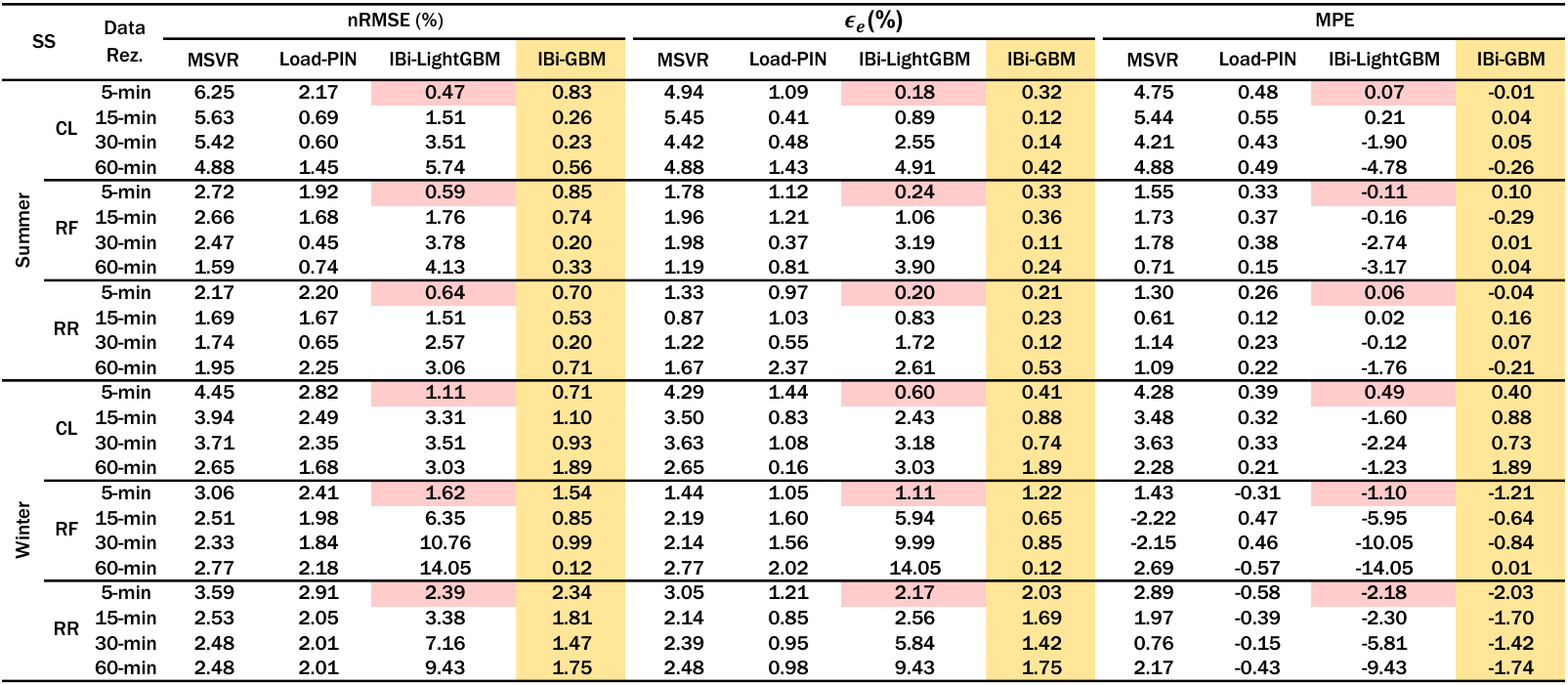}}    
\end{center}
\vspace{-0.45in}
\end{table*}

\begin{figure*}[hbt]
    \vspace{-.1in}
    \subfloat[]{%
        \centerline{\includegraphics[width=0.925\linewidth, height=0.1\textheight]{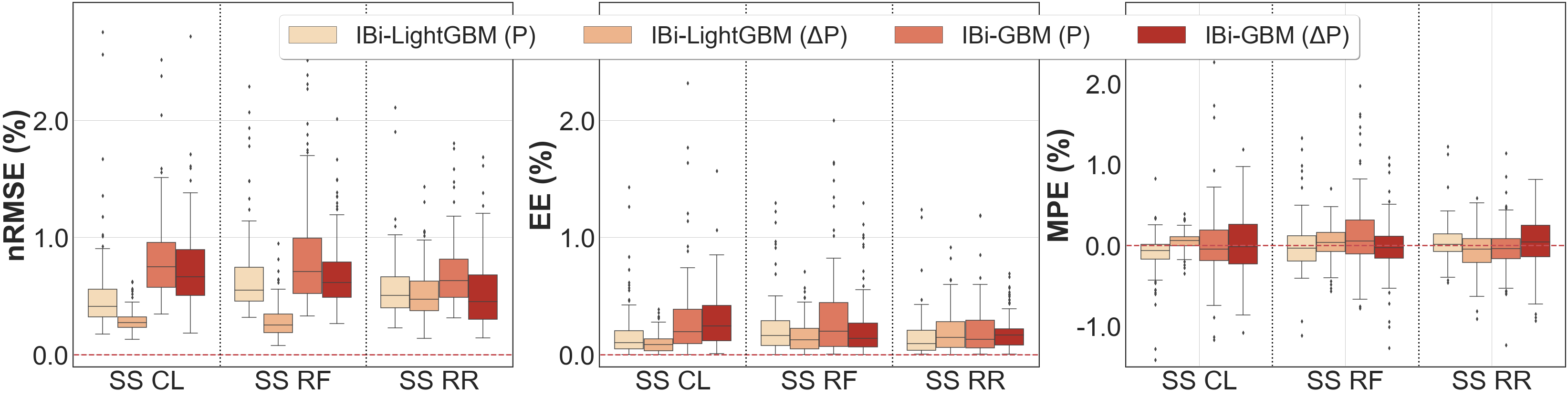}}%
        \label{11a}%
    }\hfill \vspace{-.125in} \\
    \subfloat[]{%
        \centerline{\includegraphics[width=0.925\linewidth, height=0.1\textheight]{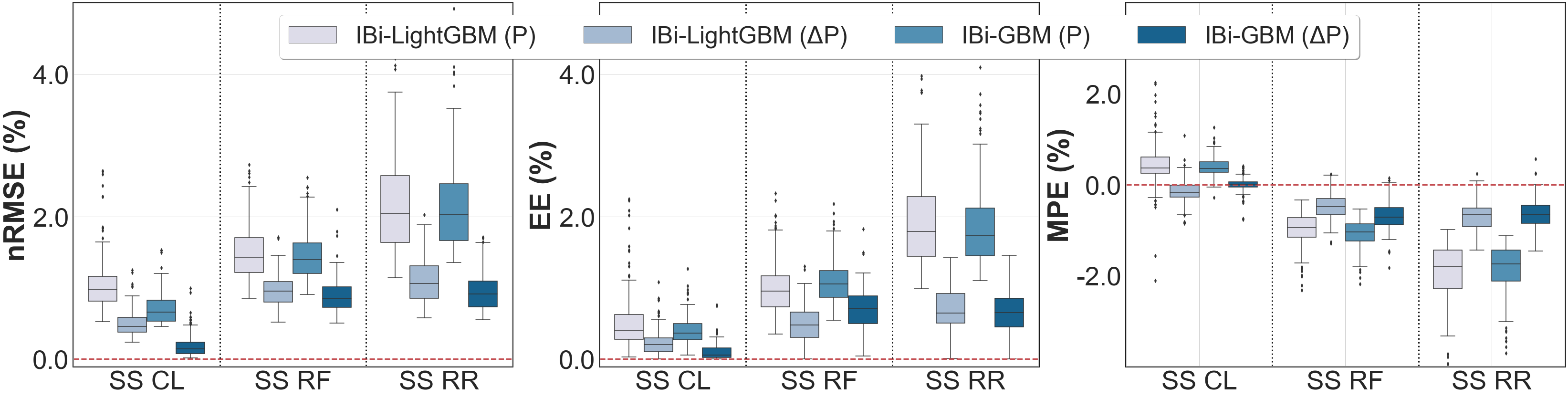}}%
        \label{11b}%
    }
    \vspace{-0.075in}    
    \caption{Prediction accuracy of two different prediction targets using IBi-GBM and IBi-LightGBM for three substations with 5-min SCADA data. (a) Summer months and (b) Winter months.}
    \label{fig11}
\vspace{-.15in}
\end{figure*}

\begin{figure}[ht]
    \vspace{-.1in}
    \subfloat[]{%
        \centerline{\includegraphics[width=0.98\linewidth, height=0.1\textheight]{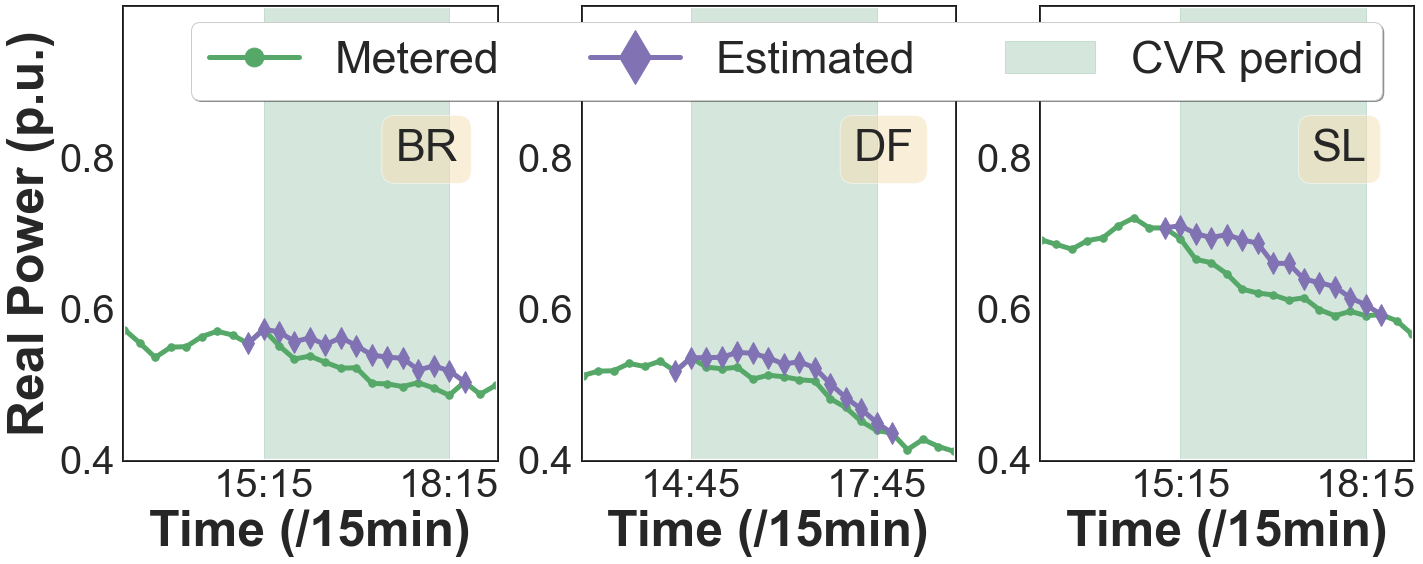}}%
        \label{12a}%
    }\hfill \vspace{-.1in} \\
    \subfloat[]{%
        \centerline{\includegraphics[width=0.98\linewidth, height=0.1\textheight]{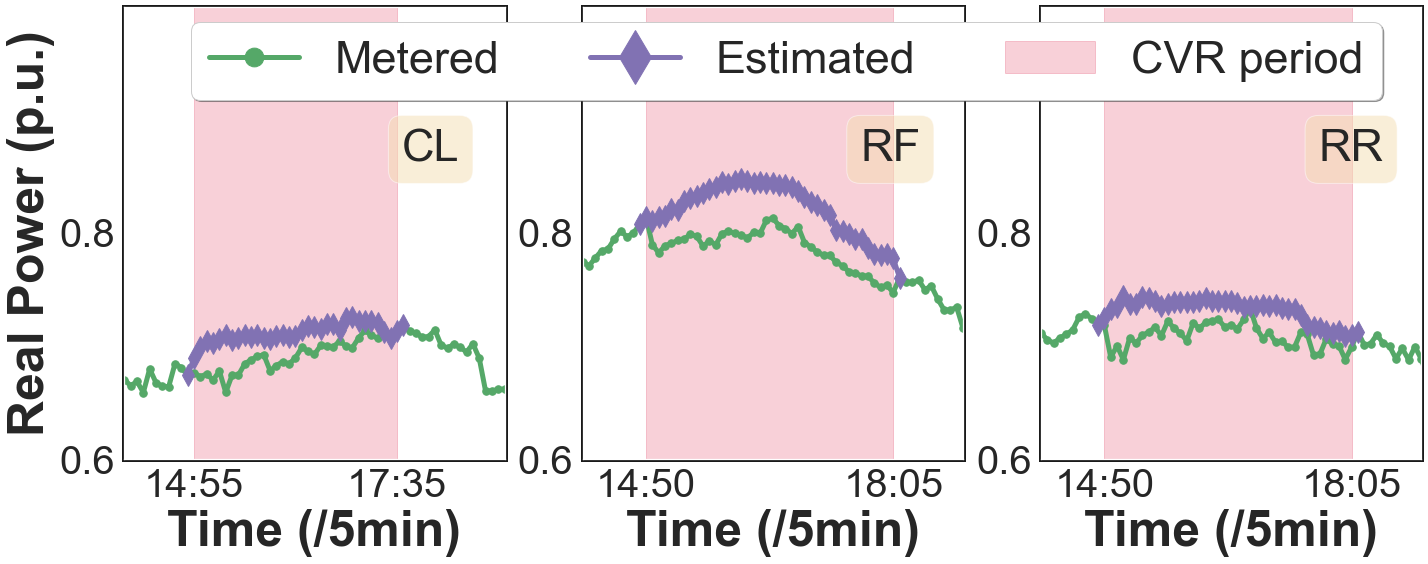}}%
        \label{12b}%
    }\hfill \vspace{-.1in} \\
    \subfloat[]{%
        \centerline{\includegraphics[width=0.98\linewidth, height=0.1\textheight]{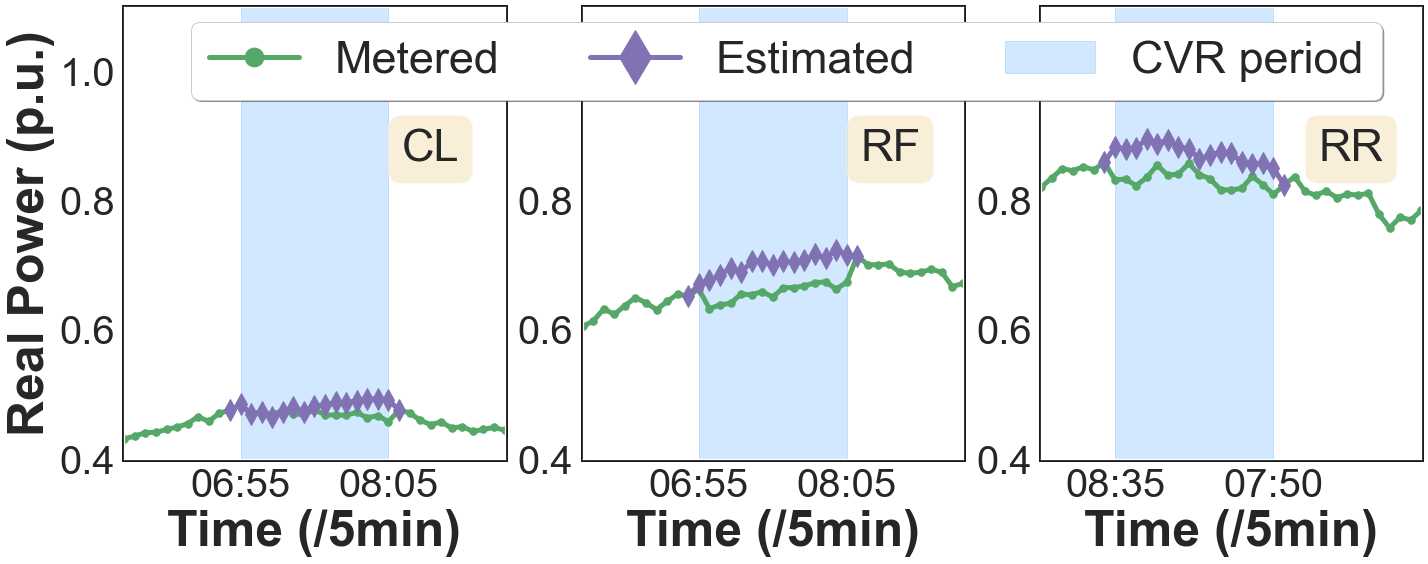}}%
        \label{12c}%
    }
    \vspace{-0.05in}    
    \caption{Examples of the IBi-GBM generated CVR baseline for (a) three feeders with 15-min smart meter data, (b) three substations with 5-min SCADA data in summer, and (c) three substations with 5-min SCADA data in winter.} 
    \label{fig12}
\vspace{-0.25in}
\end{figure}

\begin{figure}[ht]
    \vspace{-.1in}
    \subfloat[]{%
        \centerline{\includegraphics[width=0.98\linewidth, height=0.1\textheight]{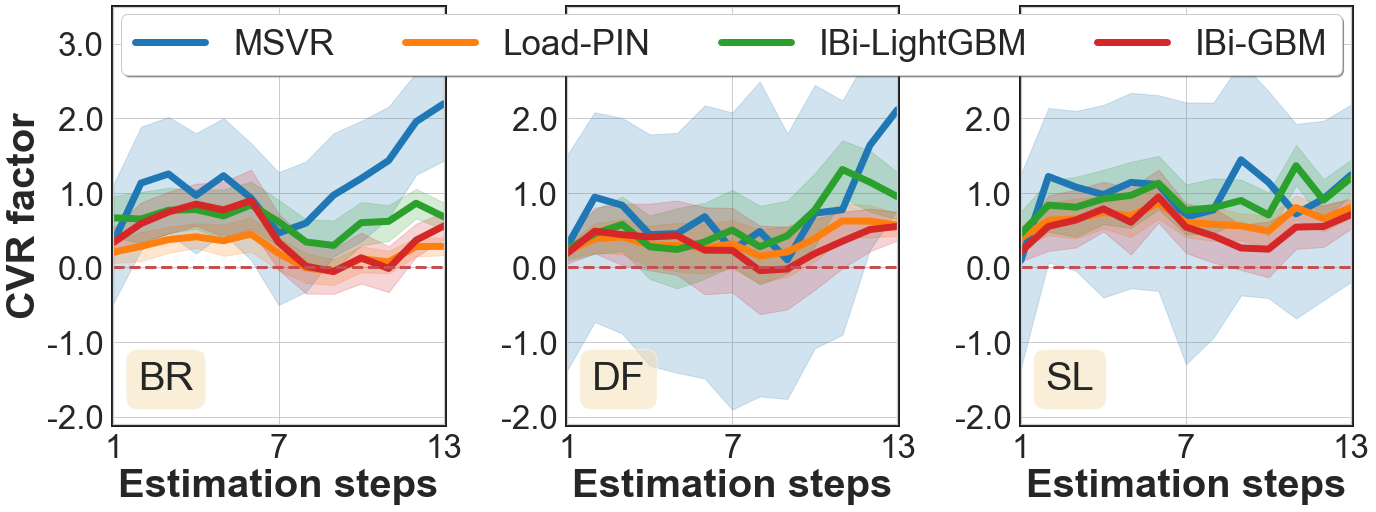}}%
        \label{13a}%
    }\hfill \vspace{-.1in} \\
    \subfloat[]{%
        \centerline{\includegraphics[width=0.98\linewidth, height=0.1\textheight]{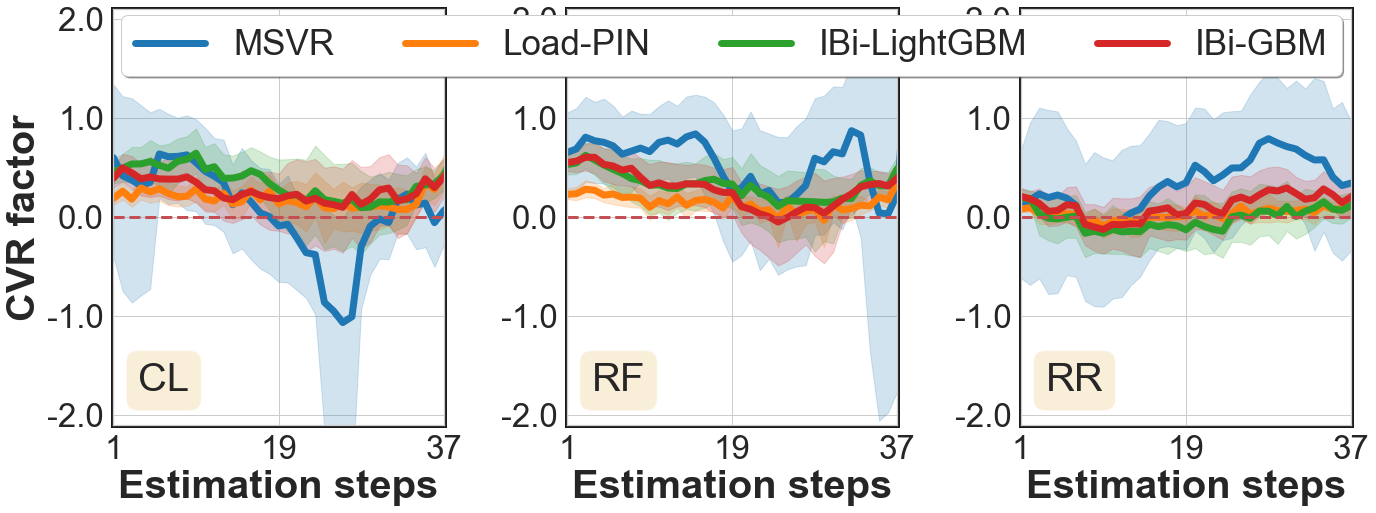}}%
        \label{13b}%
    }\hfill \vspace{-.1in} \\
    \subfloat[]{%
        \centerline{\includegraphics[width=0.98\linewidth, height=0.1\textheight]{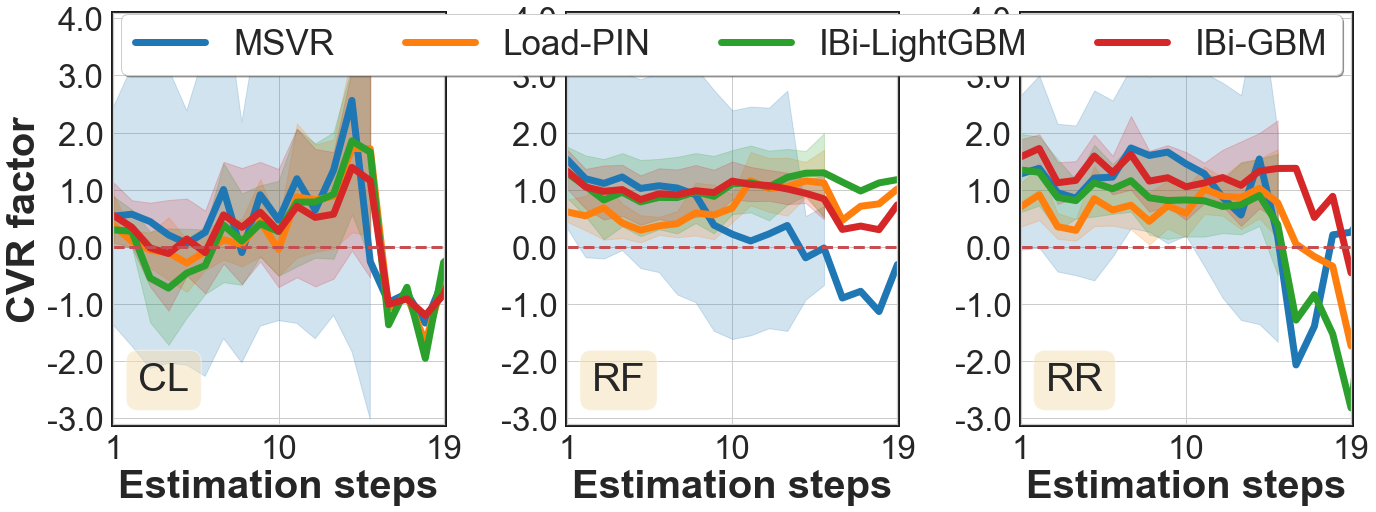}}%
        \label{13c}%
    }
    \vspace{-0.05in}    
    \caption{Average step-by-step CVR factors and 95\% confidence intervals for (a) three feeders with 15-min smart meter data, (b) three substations with 5-min SCADA data in summer, and (c) three substations with 5-min SCADA data in winter.} 
    \label{fig13}
\vspace{-.35in}
\end{figure}

\subsubsection{Uni-directional versus bi-directional forecast reconciliation}
Figure~\ref{fig8} shows the reconciliation weighting factors for three different virtual CVR durations, which are calculated using the method introduced in Section~II.E. Figures~\ref{fig8}(a) and (b) depict the outcomes of extending the CVR duration from 1-hour (yellow, 17:00-18:00) to 2-hour (orange, 16:00-18:00) and 3-hour (15:00-18:00), respectively. In addition, Figure~\ref{fig8}(c) illustrates the results of increasing the CVR duration from 30-minute (yellow, 08:00-08:30) to 1-hour (orange, 07:30-08:30) and 1.5-hour (07:00-08:30). As the weighting factor is proportional to prediction accuracy, we can observe that it is large in both directions in the beginning and then gradually diminishes towards the end. Note that, except for substations~RF and RR in Fig~\ref{fig8}(c), comparable trajectories are evident despite the extended duration. Thus, once a set of weighting factors for the maximum CVR duration is determined, it can generally be applied to other CVR durations.

Figure~\ref{fig9} compares the estimation errors (i.e. MPE by \eqref{eq13}) of the ground truth load profile (i.e., $P^{GT}$) with uni-directional forecasts (i.e., $\hat{P}^{f}$ and $\hat{P}^{b}$) and IBi-GBM forecasts (i.e., $\hat{P}^{R}$) for virtual CVR days.

As expected, the error increases as the time step shifts from 1 to end in the forward pass, and from end to 1 in the backward pass while the reconciled forecasting baseline shows much lower errors across all CVR points. The simulation results show that baseline forecasting accuracy of the forward pass is higher for the beginning few data points and the accuracy decays slower than that of the backward pass.

\subsubsection{Impact of using Iterative Reconciliation}
To demonstrate the performance improvements of using iterative reconciliation, we plot the error distributions of the reconciliation results from 1-iteration and the proposed N-iteration approaches for data from three substations (i.e., SS-CL, SS-RF, and SS-RR) in Fig.~\ref{fig10}. As can be seen in the figure, using only 1-iteration, the results has much wider distribution with a larger mean error and standard deviation, compared to the N-iteration approach.

\subsubsection{Impact for Different Data Resolution}
Next, we compare the algorithm performance when using both aggregated smart meter data and SCADA data with four data resolutions (i.e., 5-, 15-, 30-, and 60-minute). Again, the test is conducted for the virtual CVR days. Tables~\ref{tab4} and \ref{tab5} report that, except for IBi-LightGBM, the remaining models exhibit the highest accuracy at 15- and 30-minute resolutions. IBi-GBM outperforms over other prediction models across the majority of datasets. However, at the 5-minute resolution, IBi-LightGBM exhibits the highest prediction accuracy. IBi-LightGBM demonstrates strong performance in high-resolution datasets, benefiting from an abundance of training data. However, its effectiveness diminishes when applied to low-resolution datasets, where the scarcity of training data leads to underfitting and subsequently results in poorer performance.

\subsubsection{Estimation Accuracy of Different Prediction Models}
In this section, we assess the performance of the proposed methods (i.e., IBi-LightGBM and IBi-GBM) in comparison to the benchmarking methods MSVR \cite{wang2013analysis} and Load-PIN \cite{li2023load}. The Load-PIN model is trained using non-CVR days, excluding virtual CVR days, and then tested with virtual CVR days. MSVR and the proposed methods are trained using preselected similar profiles and evaluated on virtual CVR days.

In Tables~\ref{tab4} and \ref{tab5}, it can be seen that the MSVR prediction accuracy is quite low, which may be because we only used temperature profiles as weather input data and adjusted the parameter constraints used in \cite{wang2013analysis}. Furthermore, IBi-LightGBM demonstrates notably lower prediction accuracy with low-resolution data, whereas the accuracy of IBi-GBM further improves under similar conditions. This discrepancy arises because LightGBM is generally faster and more memory efficient, making it suitable for large datasets \cite{ke2017lightgbm}, while GBM performs better when dealing with small datasets or when interpretability is a priority \cite{natekin2013gradient}.

\subsubsection{Estimation Accuracy of Different Prediction Targets}
Figure~\ref{fig11} depicts the prediction accuracy of two different prediction targets using IBi-GBM and Bi-ILightGBM for three substations (i.e., CL, RF, and RR) during summer and winter, respectively, with 5-min resolution data. Note that using $\Delta P$ as the prediction target, the error of which can be bounded within a smaller range, results in lower estimation errors compared to when $P$ is used. This demonstrates that forecasting the change in load, $\Delta P$, can be more effective than directly predicting load $P$ when aiming to predict a very small \%CVR effect.

\vspace{-0.15in}
\subsection{Performance Evaluation on the \textbf{Actual} CVR Day}
All the simulations are executed on the computer with Intel core (TM) i9-10900k CPU at 3.70 GHz and 128-GB RAM. The average elapsed CPU time per CVR day of the proposed baseline estimation algorithms is 28 seconds for IBi-LightGBM and 31 seconds for IBi-GBM.

\subsubsection{CVR Performance Estimation}
After the algorithm performance is validated for the virtual CVR days, we applied the proposed algorithm on the actual CVR days. Figure~\ref{fig12} shows the examples of the IBi-GBM generated CVR baseline for the actual CVR days. The CVR factor for each time step calculated by \eqref{eq1} is shown in Fig.~\ref{fig13}. 

We made the following observations:
\begin{itemize}
    \item CVR factors are very inconsistent across feeders due to different load compositions. Feeders~BR, DF, and SL show load reduction across all 3 hours but the reduction is more prominent in the first two hours. 
    \item As illustrated in Fig.~\ref{fig13}, we observe an interesting step-response-like phenomena during the CVR event. Immediately after a CVR event is executed, the load drops due to the voltage reduction. However, after an hour or so, the load will bounce back, sometimes even higher than the baseline. After that, the load decreases again, making the overall response similar to a step response. We think that this phenomenon is caused by air conditioning loads, which require fixed amounts of energy to cool buildings. At lower voltage, air conditioners turn on longer, causing the aggregated load to bounce back after the initial drop.    
    \item Substation~RR does not show load reduction in summer. Instead, the load increases in hour 2. We think this is because the proportion of residential load on substation~RR is much higher than that of substations~CL and RF, which are predominately commercial and industrial loads. As we explained before, because the cooling load requires a fixed amount of energy to cool the house in late afternoon hours, reducing instantaneous power consumption will not reduce energy consumption. Because approximately 80\% of the residential electricity consumption is cooling loads in a hot summer afternoon, executing CVR may not reduce hourly electricity consumption. Due to increased current and increasing cooling needs in the late afternoon, the energy consumption can even increase slightly compared with the baseline.
    \item As shown in Fig.~\ref{fig13}(c), there were 7 CVR days tested in winter, out of which only 1 day lasted longer than 80-min. The CVR factor during winter is significantly higher than that in summer, possibly attributed to the substantial resistive heating load in winter.  
\end{itemize}

\subsubsection{CVR Efficacy Analysis}
\begin{figure}[t!]
    \vspace{-.1in}
    \subfloat[Substation RF]{%
        \centerline{\includegraphics[width=0.925\linewidth, height=0.1\textheight]{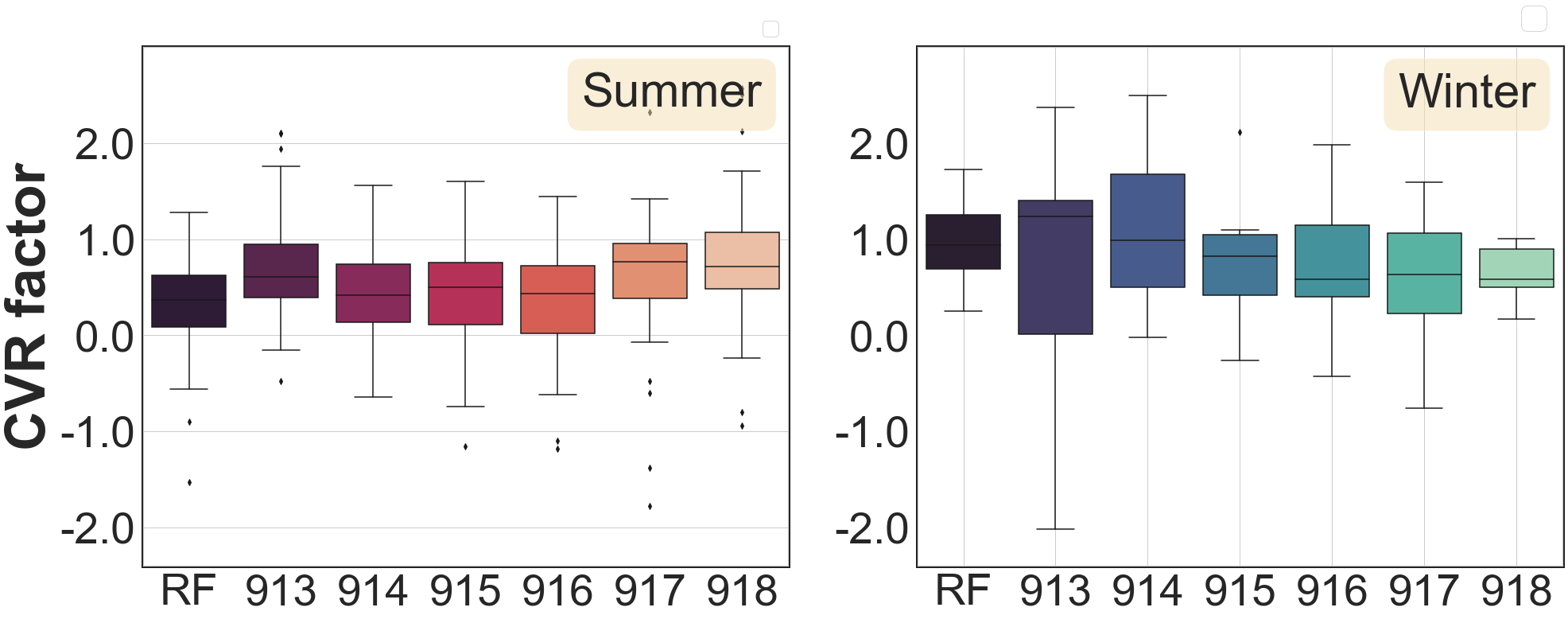}}%
        \label{14a}%
    }\hfill \vspace{-.15in} \\
    \subfloat[Substation RR]{%
        \centerline{\includegraphics[width=0.925\linewidth, height=0.1\textheight]{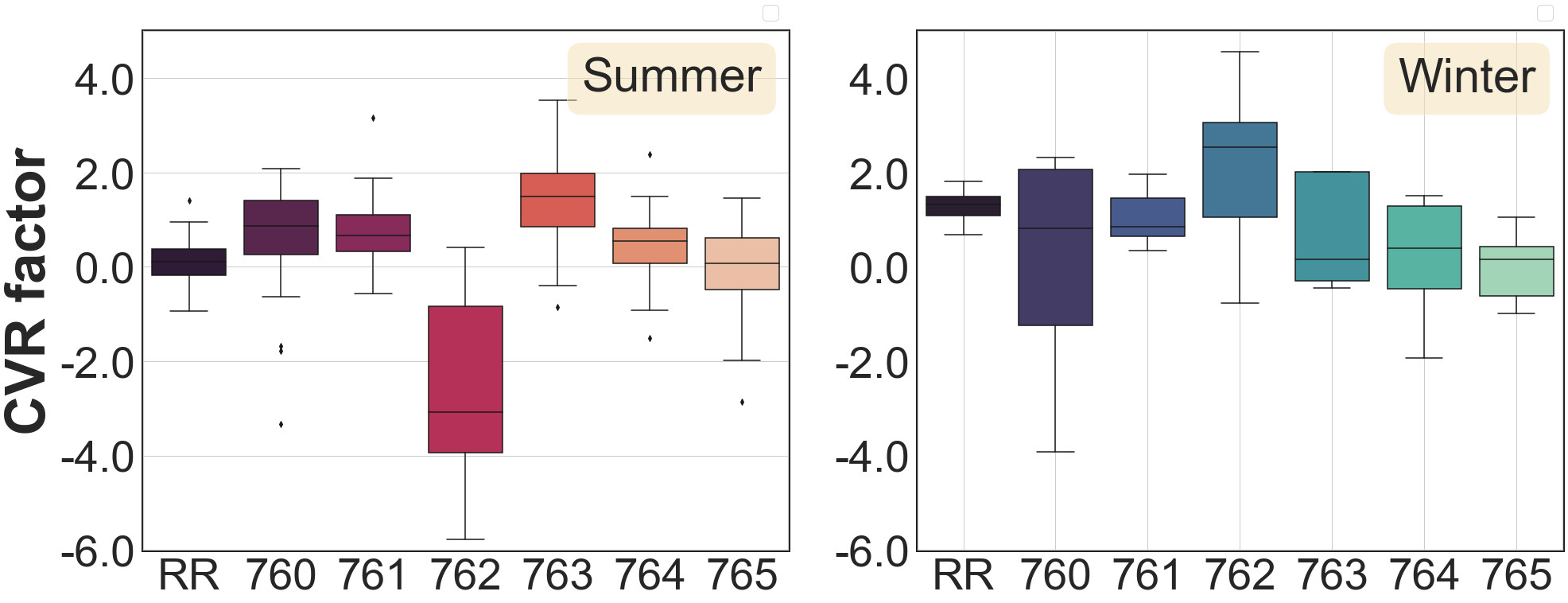}}%
        \label{14b}%
    }
    \vspace{-0.05in}    
    \caption{Seasonal CVR factors applying IBi-GBM for two substations and for each of the six feeders under each substation using 5-min SCADA data.} 
    \label{fig14}
\vspace{-.3in}
\end{figure}

Figure~\ref{fig14} shows seasonal CVR factors for two substations and for each of the six feeders under each substation. Due to the variation in CVR effects across different feeders and seasons, utilities must carefully choose the appropriate feeder to ensure consistent and significant load reductions during target CVR periods. For substation~RR, more significant CVR effects can be achieved by excluding feeder~762 in summer and feeder~760 in winter.

\vspace{-.1in}
\section{Conclusion}
In this paper, we present IBi-GBM, a novel iterative, bidirectional gradient boosting CVR baseline estimation algorithm designed to assess the efficacy of CVR in load reduction. Our approach introduces a hybrid methodology, incorporating a bi-directional framework and a hybrid similar day selection method, contrasting with the conventional uni-directional approach and power-based similar day selection method. This enables the utilization of both pre- and post-event power and temperature data as inputs for CVR baseline estimation. The proposed algorithm exhibits robust performance across different data resolutions (ranging from 5- to 60-minute intervals), various data types (including aggregated smart meter and SCADA data), and seasonal changes (specifically, summer and winter). Notably, our findings indicate that 15- and 30-minute datasets consistently outperform datasets with 5- and 60-minute resolutions, and the CVR factor during winter surpasses that of summer. Moreover, through extensive simulations and evaluations, we investigate the optimal duration for CVR, revealing its pronounced impact during the initial stages that gradually diminishes as the process progresses toward completion. 

Our findings reveal substantial variability in CVR performance across different feeders and seasons. Given this variability, we recommend that utilities customize the selection of CVR duration and feeders by collecting both smart meter and SCADA data from similar days compared to the target CVR days. 
This approach enables cross-verification of CVR effects, enhancing prediction accuracy and facilitating the identification of suitable feeders for optimal CVR outcomes.


\ifCLASSOPTIONcaptionsoff
\newpage
\fi 
\vspace{-.1in}
\bibliographystyle{IEEEtran}
\bibliography{IEEEabrv,MyRefs}

\end{document}